\documentclass[aps,twocolumn]{revtex4}
\usepackage{epsfig}

\newcommand{\be}{\begin{equation}}
\newcommand{\ee}{\end{equation}}
\newcommand{\ba}{\begin{eqnarray}}
\newcommand{\ea}{\end{eqnarray}}

\usepackage{epsfig}
\begin{document}
\title{Dragging a polymer chain into a nanotube and subsequent release}
                                                                               
\author{Leonid I. Klushin}
\affiliation{American University of Beirut, Department of Physics,
Beirut, Lebanon}
\author{Alexander M. Skvortsov}
\affiliation{Chemical-Pharmaceutical Academy, Prof. Popova 14, 197022
St. Petersburg, Russia.}
\author{Hsiao-Ping Hsu and Kurt Binder}
\affiliation{Institut f\"ur Physik, Johannes Gutenberg-Universit\"at Mainz\\
D-55099 Mainz, Staudinger Weg 7, Germany}

\begin{abstract}
   We present a scaling theory and Monte Carlo (MC) simulation results for a 
flexible polymer chain slowly dragged by one end into a nanotube. We also 
describe the situation when the completely confined chain is released and 
gradually leaves the tube. MC simulations were performed for a self-avoiding 
lattice model with a biased chain growth algorithm, the pruned-enriched
Rosenbluth method (PERM). The nanotube is a long channel opened at one end and 
its diameter $D$ is much smaller than the size of the polymer coil in 
solution. We analyze the following characteristics as functions of the chain 
end position $x$ inside the tube: the free energy of confinement, the average 
end-to-end distance, the average number of segments imprisoned in the tube, 
and the average stretching of the confined part of the chain for various 
values of $D$ and for the number of repeat units in the chain, $N$. 
We show that when the chain end is dragged by a certain critical distance 
$x^*$ into the tube, the polymer undergoes a first-order phase transition 
whereby the remaining free tail is abruptly sucked into the tube. 
This is accompanied by jumps in the average size, the number of imprisoned 
segments, and in the average stretching parameter.  
The critical distance scales as $x^*\sim ND^{1-1/\nu}$. 
The transition takes place when approximately $3/4$ of the chain units 
are dragged into the tube. The theory presented is based on constructing 
the Landau free energy as a function of an order parameter that provides 
a complete description of equilibrium and metastable states. We argue that 
if the trapped chain is released with all monomers allowed to fluctuate, 
the reverse process in which the chain leaves the confinement occurs 
smoothly without any jumps. Finally, we apply the theory to estimate
the lifetime of confined DNA in metastable states in nanotubes.
 
\end{abstract}

\maketitle

\section{Introduction}

Recent developments in fabrication of nanoscale devices and in 
single-chain manipulation techniques open possibilities for a broad 
range of applications in biotechnology and materials 
science~\cite{Salman, Meller, Kasianowicz, Aktson, Meller00, Clausen, Williams}. 
In particular, well calibrated nanochannels were produced in fused 
silica substrates by lithography methods with the widths in 
the range of $30$ to $400$ nm, which were used to study the 
confinement of single $\lambda$-phage DNA molecules driven 
electrophoretically into these nanochannels~\cite{Reisner}. 
The persistence length of DNA under conditions used in these experiments 
is about $50$ nm while its contour length was about $1000$ times larger. 
This means that except for the case of the narrowest channels DNA behaved 
essentially as a long flexible macromolecule on the relevant length scales. 
The aim of this paper is to elucidate the subtle physics behind the process 
when a long flexible chain is slowly dragged by one end into a nanochannel. 
We also show that this process is qualitatively different from what happens 
when a confined chain is released and leaves the nanochannel by spontaneous 
thermal motion.

\begin{figure*}
\begin{center}
\epsfig{file=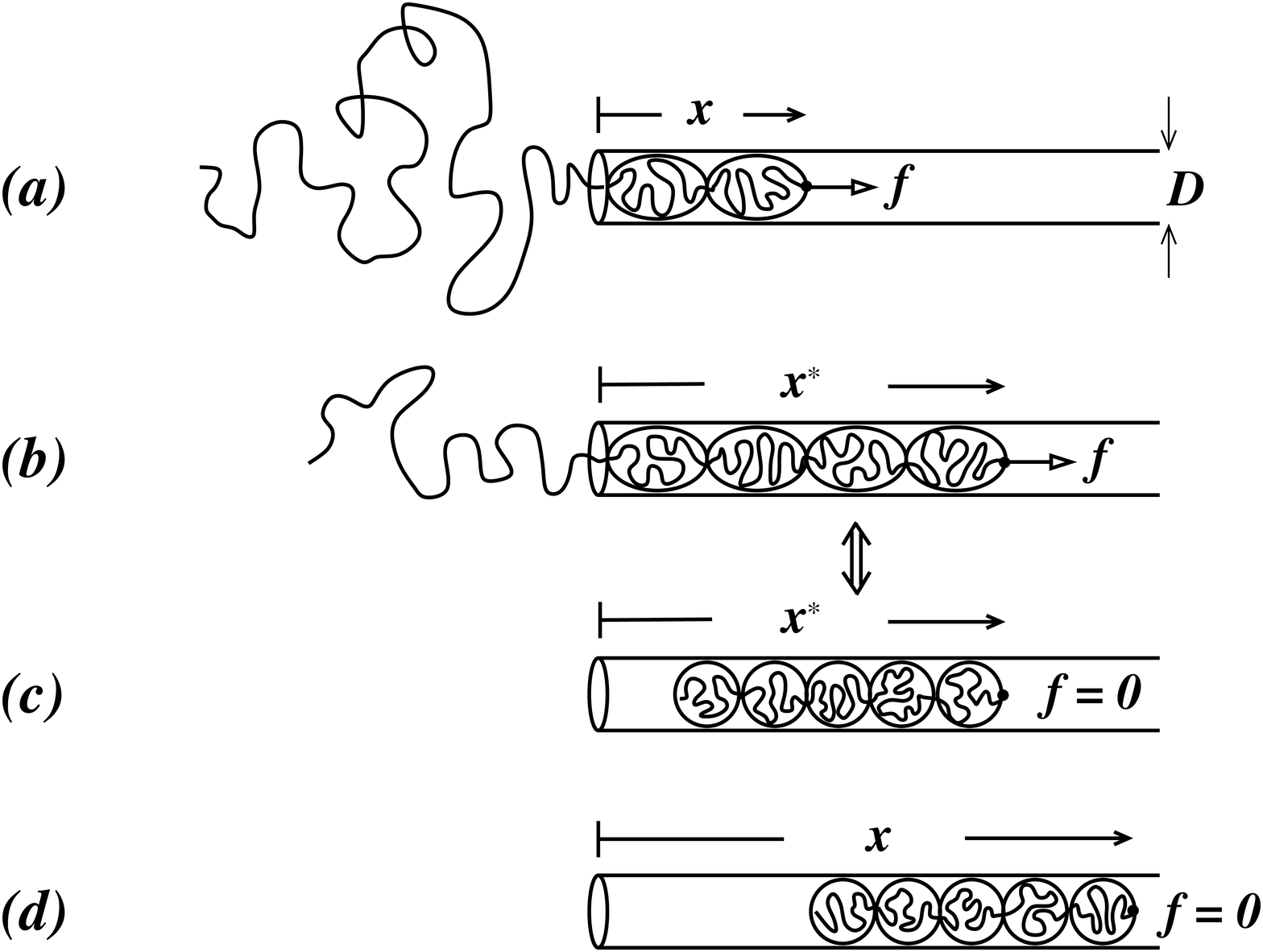, width=12.0cm, angle=0}
\caption{Schematic drawings of a flexible polymer chain with one
end dragged into a nanotube in a quasi-equilibrium process.
The fixed chain end characterized by its coordinate $x$ experiences
an  reaction force $f$ that balances the pull of the undeformed
swollen coil outside the tube (a)(b). At the transition point
$x^*$, the remaining tail is sucked into the tube abruptly by a uniform
shrinking of all blobs in the stem (b), (c), and the
reaction force becomes zero, $f=0$. At the transition point
conformations (b) and (c) coexist. As long as the chain is fully confined in the tube,
no reaction force appears at the fixed chain end (d).}
\label{drag-fig1}
\end{center}
\end{figure*}

Our approach is based on using the most general results of the scaling 
theory that neglect the small-scale details of the system under consideration, 
namely the particular form of the interaction potentials, the flexibility 
mechanisms, etc. It is well known that the scaling approach does not allow to 
calculate non-universal numerical coefficients that are model-dependent. 
To verify the prediction of the analytical theory we have carried out 
detailed Monte Carlo (MC) simulation by using the pruned-enriched Rosenbluth
method (PERM)~\cite{g97,Hsu03,Hsu04,Hsu07}. 
The simulations are based on the simplest model of polymers, namely
self-avoiding walks on a cubic lattice. Fortunately, the general ideas 
of scaling guarantee a universal behavior within broad limits of parameters.

The properties of a single macromolecule confined in a tube have been 
studied extensively for decades, both by analytical theory and by numerical 
simulations for various models of flexible and semi-flexible 
chains~\cite{Kremer, Sotta, Milchev94, Yang}.
For a homogeneous confined state there are scaling predictions~\cite{Daoud} 
concerning various chain characteristics which were tested by MC simulations. 
Our main interest here is in the non-homogeneous flower-like states where 
the confined part of the chain inside the tube forms a stretched stem and 
the free tail still in solution forms a coiled crown. This type of 
conformations appears in a variety of situations including translocation 
through a thick membrane~\cite{Randel} as well as the
escape transition produced 
by compressing a grafted chain between flat 
pistons~\cite{Subra, Ennis99, Sevick99, Steels, Milchev}. One of the 
goals of this paper is to demonstrate that the role of these conformations 
in the situations mentioned above are quite different depending on whether 
one of the chain ends is fixed in the confinement region or not.

The paper is organized as follows: We start by presenting the main 
qualitative results in a series of simple pictures visualizing the 
conformational changes in the process of dragging the chain end slowly into 
the tube and its de-confinement upon subsequent release. In Section III we 
give the results of the MC simulations for completely confined states and 
compare them with the well-known scaling prediction. Section IV describes 
the simulation results characterizing the phase transition induced by 
changing the chain end position inside the tube. In Section V 
we provide a theoretical description of the phenomenon based on the Landau 
free energy approach and compare it to the MC results. The process of 
spontaneous de-confinement is analyzed in Section VI and followed by a 
general discussion. 

\begin{figure*}
\begin{center}
\epsfig{file=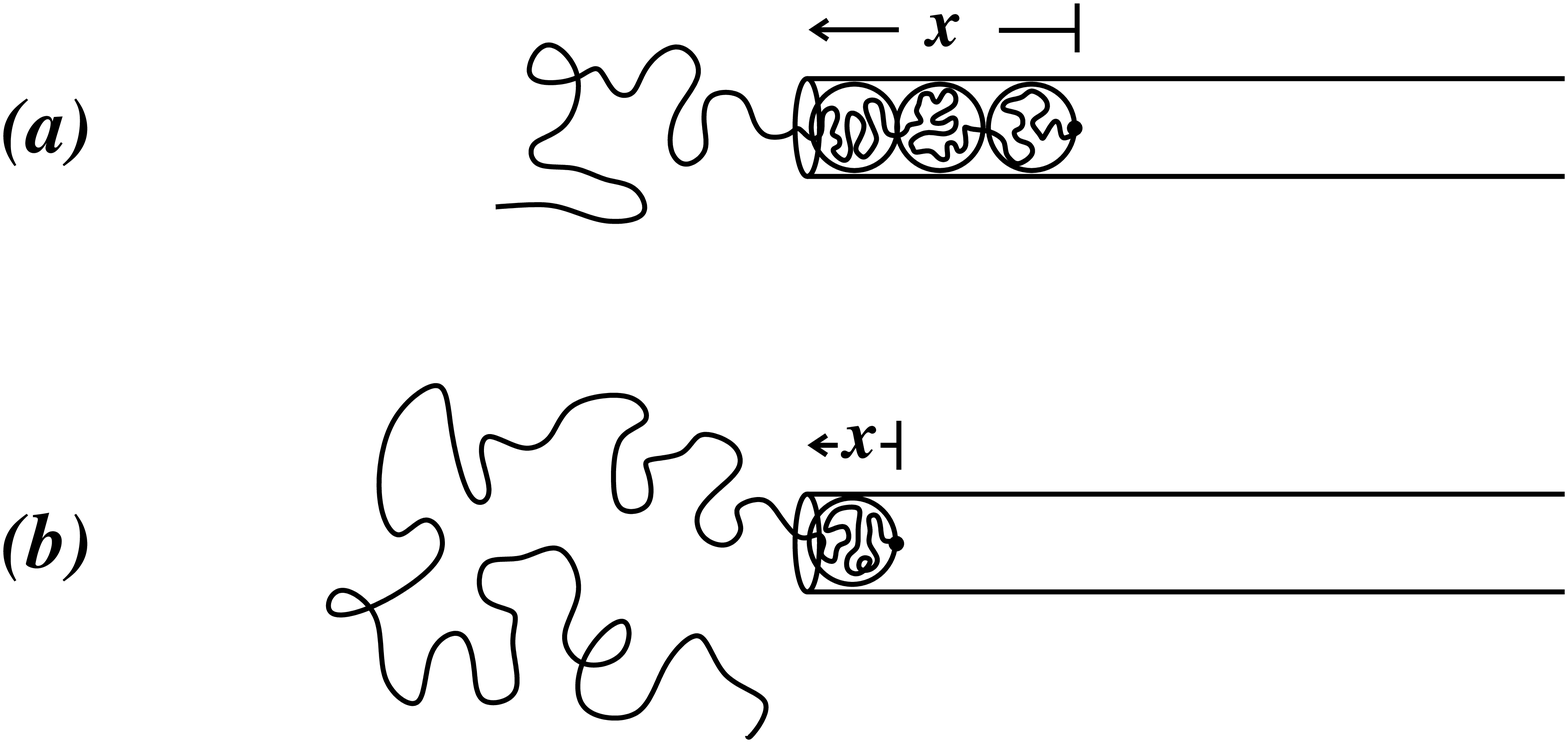, width=12.0cm, angle=0}
\caption{Schematic drawings of a released flexible polymer chain.
Since all the chain units are free to fluctuate, no reaction force
is needed to counterbalance the force due to the tail outside the tube.
The stretching of the stem is only dependent on the confinement effect
at any moment, (a) and (b).}
\label{drag-fig2}
\end{center}
\end{figure*}

\section{Qualitative picture}

Figure~\ref{drag-fig1} presents a sequence of chain conformations when the position 
of one of the chain ends is progressively moved quasi-statically further 
inside the nanotube. At each moment, this fixed chain end 
is characterized by its coordinate $x$ which is controlled externally. 
The tail outside the tube is a practically undeformed swollen coil while 
the part of the chain inside forms a one-dimensional string of blobs. The chain
can be presented as a flower structure with one-dimensional structure of the
stem. 
The number of confined (imprisoned) chain units grows linearly 
with $x$, at least in the initial stages of the process. 
The fixed chain end experiences a force which is due to the tail 
still not being confined. To counterbalance this force, a reaction force $f$ directed 
into the tube will appear. Once the chain is confined completely 
(Figures~\ref{drag-fig1}c and \ref{drag-fig1}d) there is no tail outside the tube 
and the reaction force at the 
controlled chain end disappears. 
In this state, the chain is homogeneously stretched and its stretching degree 
is due solely to the confinement effect~\cite{Daoud}. It is clear that the stem of the 
partially confined conformation is stretched more strongly since the 
confinement effects are augmented by the additional stretching force 
(this effect is symbolically indicated by the deformation in blob shape
shown in the Figures~\ref{drag-fig1}a and \ref{drag-fig1}b). 
The difference in the deformation free energy 
leads to a phase transition - an abrupt ``slurping'' of the remaining 
tail accompanied by a uniform shrinking of all blobs in the stem. 
(Of course, in a strict sense true phase transitions can occur only in
the thermodynamic limit, which would require that both the length
of the chain and the length of the pore are infinitely large;
however, as we shall see, the rounding of the phase transition caused
by finite chain length is not too significant.)
As a result the length along the tube of the completely confined chain is 
less than the length of the stem just before the transition, 
as illustrated in Figures~\ref{drag-fig1}b and \ref{drag-fig1}c.
Note that the process described above has to be distinguished from 
the case when the chain is driven into the pore by applying a constant
force, as in electrophoresis. 

The next figure (Figure~\ref{drag-fig2}) illustrates the de-confinement of a released chain. 
All the chain units are free to fluctuate and there is no reaction force. 
Therefore, the stem stretching is only due to the confinement effect at 
any particular moment. The gradual decrease in the number of imprisoned 
units is not accompanied by any jumps.

\section{Fully confined chain: MC results and scaling}

The scaling picture of a fully confined chain inside a tube of 
diameter $D$ is very simple. It is formed by a string of non-overlapping 
blobs of size $\sim D$, each blob containing $g$ monomer units~\cite{Daoud}. 
The size of the blob is related to $g$ by
\be
     D \sim ag^{\nu}    \label{blob size}
\ee
where $a$ is the length of a monomer unit, and $\nu=0.58765(20)$ is the scaling 
index for a three-dimensional self-avoiding chain~\cite{Hsu04}. 
Let's define the number of blobs by relation 
\be
     n_b=N/g = N(D/a)^{-1/\nu}   \; .  \label{blob number}
\ee
In this case the average end-to-end distance of the chain in a fully confined
(imprisoned) state is 
\be
     R_{\rm imp}=A_{\rm imp}Dn_b       \label{Rimprison}
\ee
where $A_{\rm imp}$ is a model-dependent numerical coefficient.
In eqs~\ref{blob size}-\ref{Rimprison} we have assumed that the 
pore diameter $D$ is large enough, so that the number of 
monomers inside a single blob is large enough so that the scaling
relation eq~\ref{blob size} holds; further more correction
terms to the scaling relations are omitted throughout.

\begin{figure}
\begin{center}
\includegraphics[scale=0.31,angle=270]{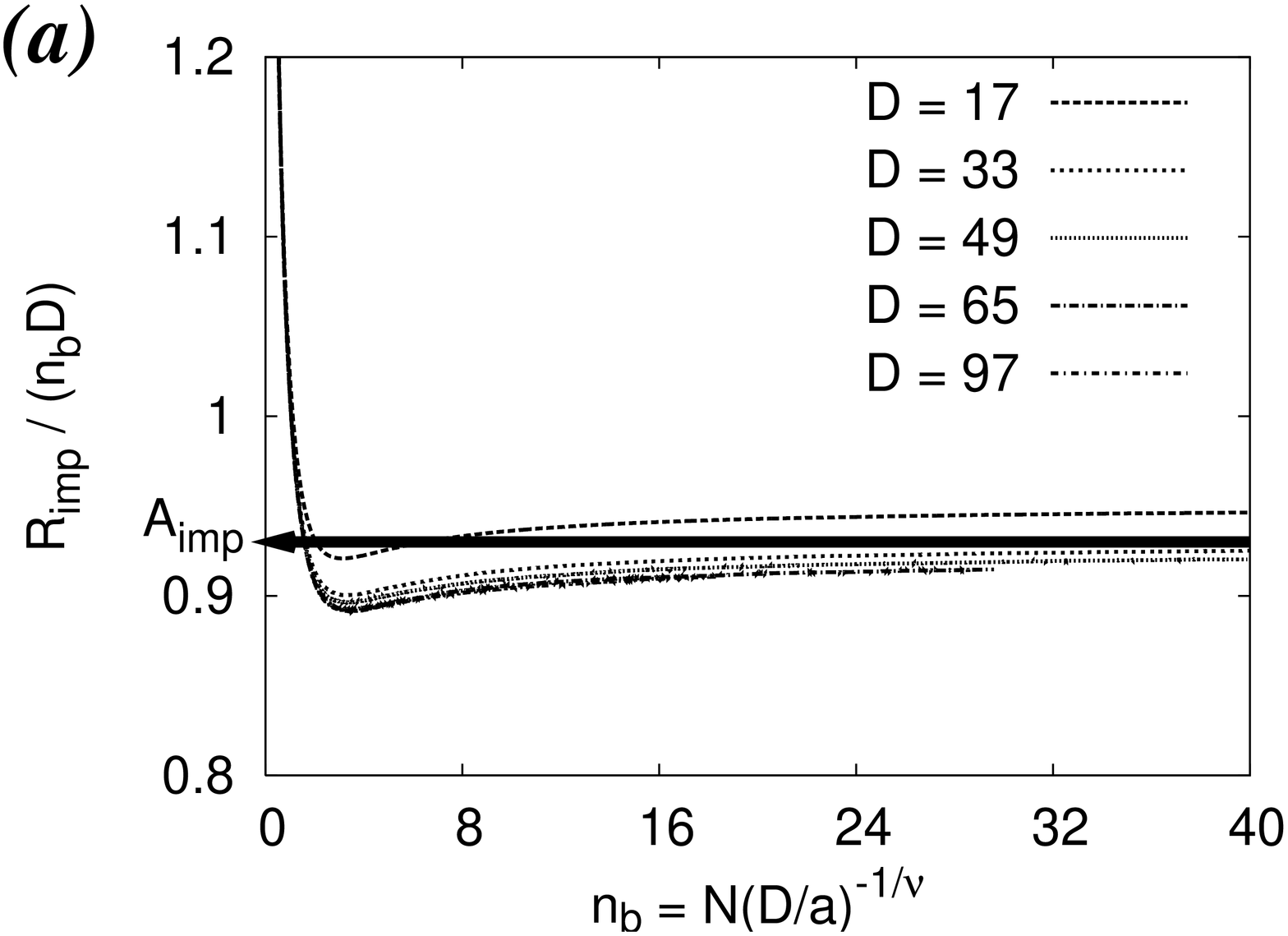}\hspace{0.4cm}
\includegraphics[scale=0.31,angle=270]{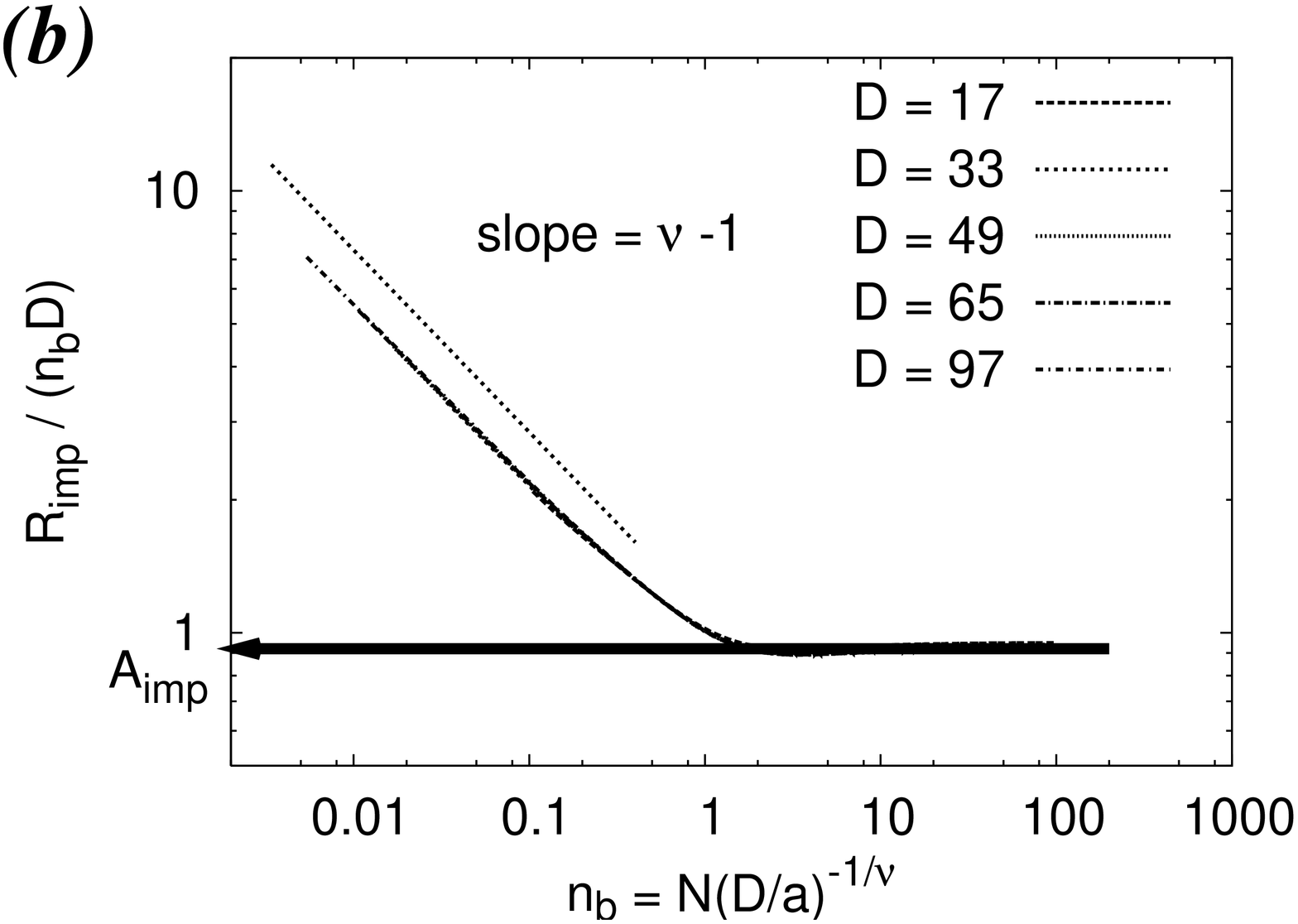}
\caption{(a) Rescaled average end-to-end distance $R_{\rm imp}/(n_b D)$
against the number of blobs $n_b=N(D/a)^{-1/\nu}$ for fully confined chains.
The thick solid line with arrow is $R_{\rm imp}/(n_b D)=A_{\rm imp}$
 with $A_{\rm imp}=0.92(3)$. (b) The log-log plot of the same data as
(a) shows that for $n_b>2$ the data collapse  and allows to
estimate $A_{\rm imp}$ very accurately.
The dotted straight line with slope $\nu-1$ shows
the scaling law in the regime of wide tubes with $D > R_{\rm imp}$.}
\label{drag-fig3}
\end{center}
\end{figure}

\begin{figure}
\begin{center}
\includegraphics[scale=0.31,angle=270]{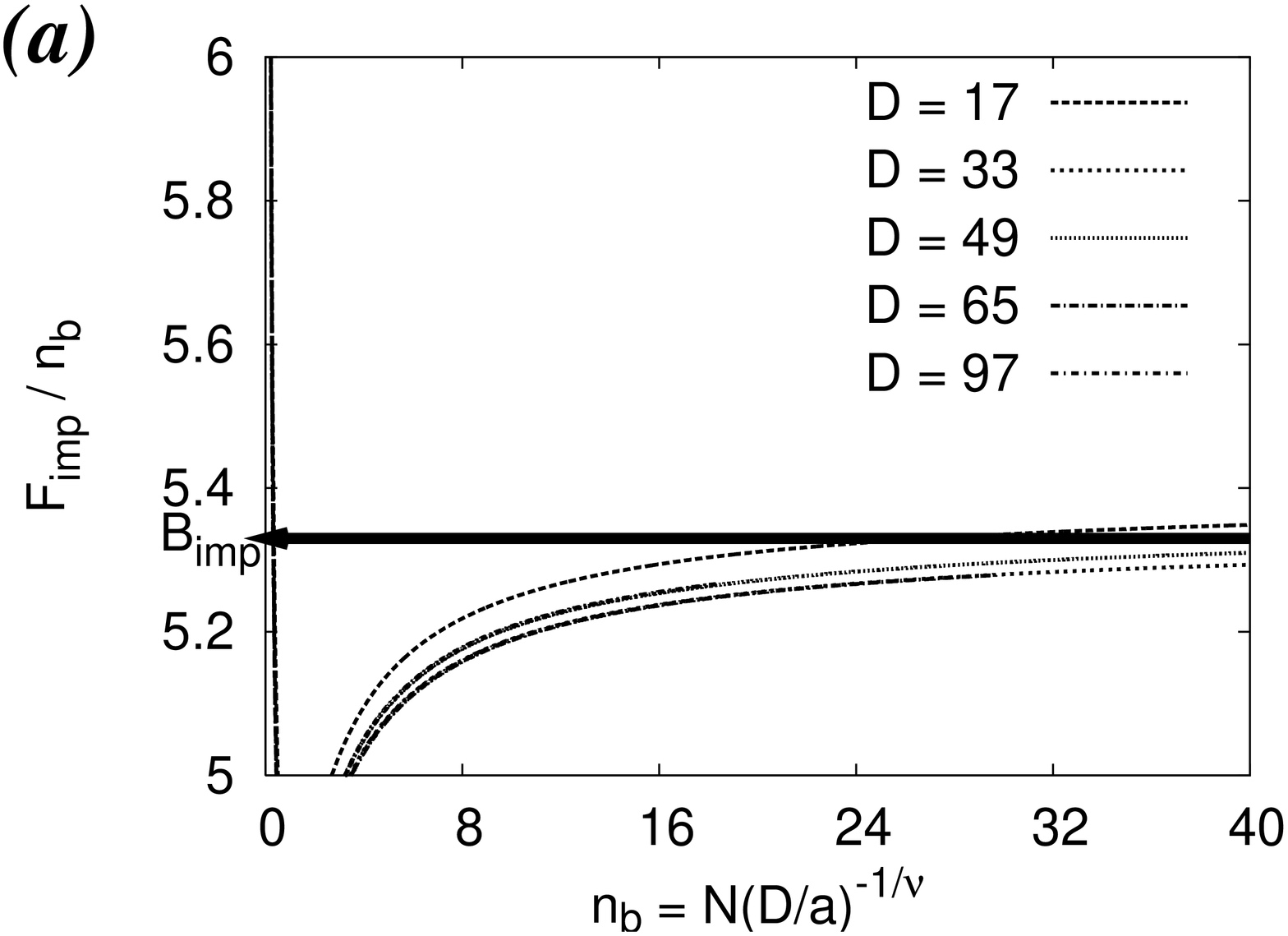}\hspace{0.4cm}
\includegraphics[scale=0.31,angle=270]{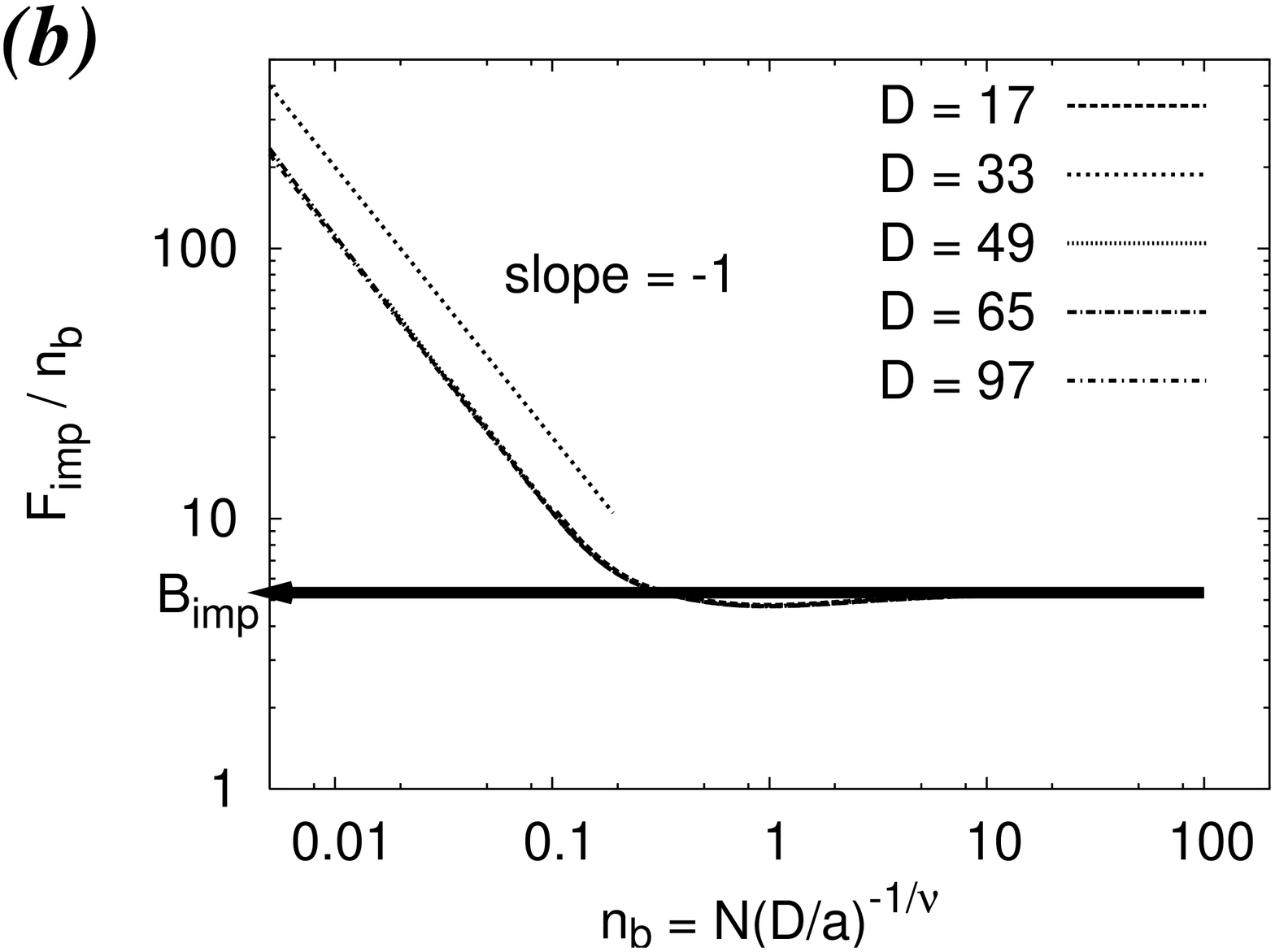}
\caption{(a) Rescaled average free energy $F_{\rm imp}/n_b$ for fully
confined chains against the number of blobs $n_b$ .
The thick solid line with arrow is $F_{\rm imp}/n_b=B_{\rm imp}$
 with $B_{\rm imp}=5.33(5)$. (b) Same as (a), but on a
log-log plot to show that for $n_b>1$ the data allows to estimate $B_{\rm imp}$
very accurately.}
\label{drag-fig4}
\end{center}
\end{figure}

A stretching parameter $S_{\rm imp}=R_{\rm imp}/Na$  describes the average 
stretching of the chain in an imprisoned state. Since the end-to-end
distance of the fully imprisoned chain is proportional to $N$, 
the stretching parameter is a function of the tube diameter only and 
is given by 
\be
     S_{\rm imp}=A_{\rm imp}(D/a)^{1-1/\nu} \;.    \label{Simprison}
\ee
For a fully imprisoned chain the free energy of confinement per blob 
scales as $k_BT$. Thus, using eq~\ref{blob number} the free energy of a fully
imprisoned chain is 
\be
     F_{\rm imp}=B_{\rm imp} n_b    \label{Fimprison}
\ee
where $B_{\rm imp}$ is another numerical coefficient. The factor of $k_BT$
is absorbed in the free energy throughout the paper hereafter.

For our simulations, chain lengths are up to 44000, tube diameters are up to $D=97$,
and $a=1$ which is the lattice spacing.
Simulation data for the rescaled average root mean square (rms) end-to-end
distance $R_{\rm imp}/(n_bD)$ are displayed in Figure~\ref{drag-fig3} 
depending on the blob number $n_b$. 
It is evident from the Figure~\ref{drag-fig3}
that the normalized chain size reaches a practically constant value when the
chain contains more than two blobs. The limiting constant value of $R_{\rm
imp}/(n_bD)$ is just the numerical coefficient $A_{\rm imp}=0.92\pm 0.03$ in
eq~\ref{Rimprison}. As noted above, the scaling description is
expected to become exact in the asymptotic limit $D \rightarrow \infty$,
and hence in the considered range of not very large $D$ a small
systematic dependence of the curves in Figure~\ref{drag-fig3}a
is evident, leading to the spread of values of the coefficient $A_{\rm imp}$,
as expressed by the quoted uncertainty.  
Figure~\ref{drag-fig3} shows that at values $n_b<1$, the unperturbed coil
size smaller than the tube diameter, there is another scaling regime of weak
confinement where $R_{\rm imp} \sim N^{\nu}$. Although the crossover between these
two regimes is of general theoretical interest, we are not concerned with it in
this paper. 

The rescaled free energy $F_{\rm imp}$ (counted from the reference state of a
self-avoiding coil in free space) for the fully imprisoned chain is presented in
Figure~\ref{drag-fig4} vs. the number of blobs $n_b$. The limiting constant value
gives the coefficient $B_{\rm imp}=5.33 \pm 0.08$ in eq~\ref{Fimprison}, which
represents the free energy per blob in $k_BT$ units.
Once again we see the other
scaling regime of weak confinement if the number of blobs in the chain is less then
one blob. Summarizing the presented data we can conclude that using the definition
of the number of blobs given by eq~\ref{blob number} the size of
the blob in our model is close to $D$, the end-to-end distance
$R_{\rm imp} \sim 0.92n_bD$, and the free energy per blob
$F_{\rm imp}/n_b \sim 5.33$ is close to 5 $k_BT$. Comparing our result
for the free energy per blob, $F_{\rm tube}/n_b$ $(F_{\rm tube}=F_{\rm imp})$
with that of chains confined in a slit
on a lattice~\cite{Hsu04}, $F_{\rm slit}/n_b \sim 2.10$,
and a off-lattice model~\cite{Dimitar}, $F_{\rm slit}/n_b \sim 2.03$, we find
that $F_{\rm tube} \approx 2 F_{\rm slit}$. It is not
surprising because a polymer chain confined in a slit is compressed in one direction
and in a tube in two directions.

\begin{figure}
\begin{center}
\epsfig{file=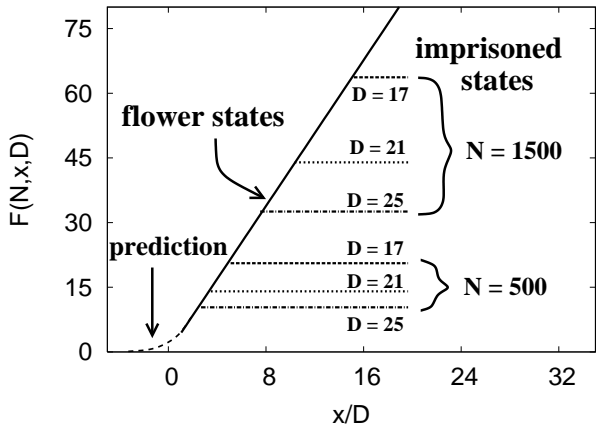, width=6.0cm, angle=270}
\caption{Free energy of chains with chain lengths $N=500$ and $N=1500$
relative to a self-avoiding
coil, $F(N,x,D)$, plotted against $x/D$ for tube diameters $D=17$, $21$ and $25$.
The solid straight line is $F_{\rm fl}=4.23x/D$ and gives the best fit of the
data of chains in a flower state. All horizontal lines indicate the values of
$F_{\rm imp}$ for chains in an imprisoned state at fixed chain length $N$. The
intersections of the solid line and horizontal lines indicate the transition
points $x^*/D$.}
\label{drag-fig5}
\end{center}
\end{figure}

\begin{figure}
\begin{center}
\epsfig{file=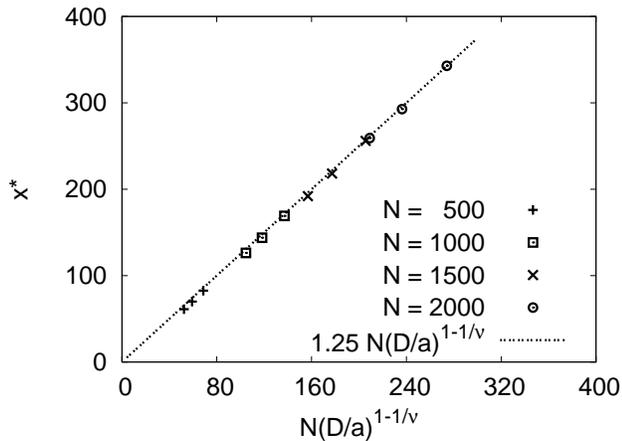, width=6.0cm, angle=270}
\caption{Transition points $x^*$ as obtained from Figures~\ref{drag-fig5},
plotted against $N(D/a)^{1-1/\nu}$ for various values of $N$ and $D$.
The dotted line has slope $B_{\rm imp}/B_{\rm fl}=1.25$ .}
\label{drag-fig6}
\end{center}
\end{figure}

\section{Phase transition: equilibrium characteristics}

For our simulations, single polymer chains are dragged into
a tube with diameters $D=17$, $21$, $25$, and $29$, and the chain
length is up to $N=17000$.
Simulation results of the free energy relative to a self-avoiding coil
are presented in Figure~\ref{drag-fig5} as a function of the end monomer
position inside the tube, $x$, normalized by the tube diameter $D$.
It is clear that there are two branches of the free energy.
Initially, the free energy increases linearly with $x$ as more and more
blobs are driven into the tube. The slope of the free energy as a function
of $x$ has the meaning of the average reaction force acting on the
end monomer. Deviations from the linearity near the origin occur when
only the number of blobs inside the tube is of order one or less. At large
enough values of $x$ all monomeric units are confined and the free energy does not
depend on the position $x$ any more. An abrupt change in the slope of the
free energy indicates a first-order transition.
The linear branch describes a partially confined ``flower'' state.
On this branch the data points for different values of $N$ and $D$ collapse onto the
same universal curve in the chosen coordinates. The $x$-independent branch
corresponds to a completely confined state discussed above in Section III.
Altogether the results can be summarized as follows:
\be
F(N,x,D) = \left\{
\begin{array}{ll}
B_{\rm imp}  N(D/a)^{-1/\nu}=F_{\rm imp} & \enspace {\rm imprisoned \, state} \\
 B_{ \rm fl} ( x/D)=F_{\rm fl}  &\enspace {\rm flower \, state}
\end{array} \right .   \; .
\label{Fscaling}
\ee
where $B_{\rm imp}= 5.33(8)$ and $B_{\rm fl} = 4.23(6)$ were obtained
from Figure~\ref{drag-fig4} and Figure~\ref{drag-fig5}. Physically, both
formulas state that the free energy is proportional to
the number of blobs inside the tube. However, we would like to point out
that the numerical coefficients are different, and the most important source
of this difference is due to the extra stretching of the flower stem as
compared to the relaxed fully confined state.

\begin{figure}
\begin{center}
\includegraphics[scale=0.31,angle=270]{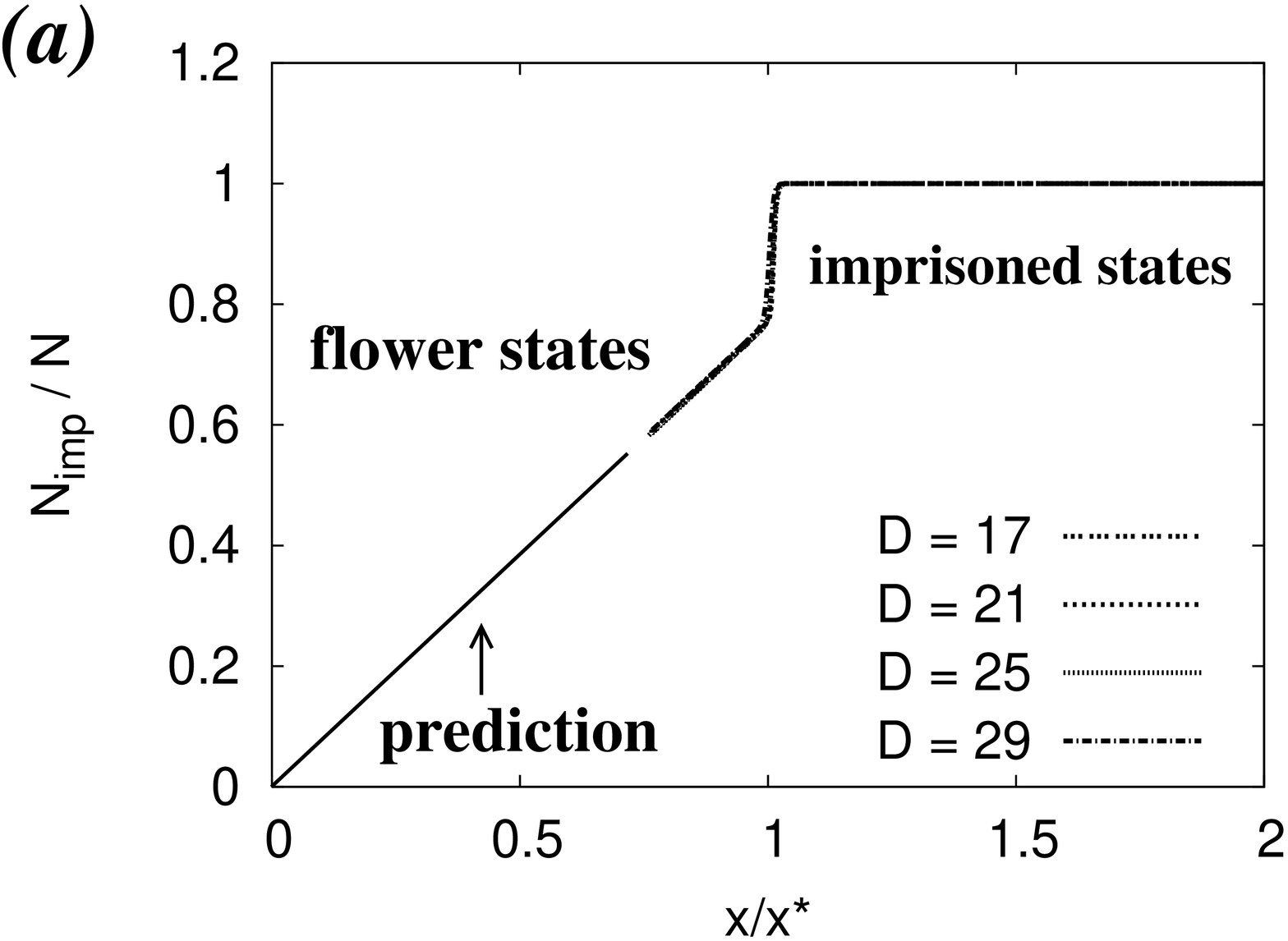}\hspace{0.4cm}
\includegraphics[scale=0.31,angle=270]{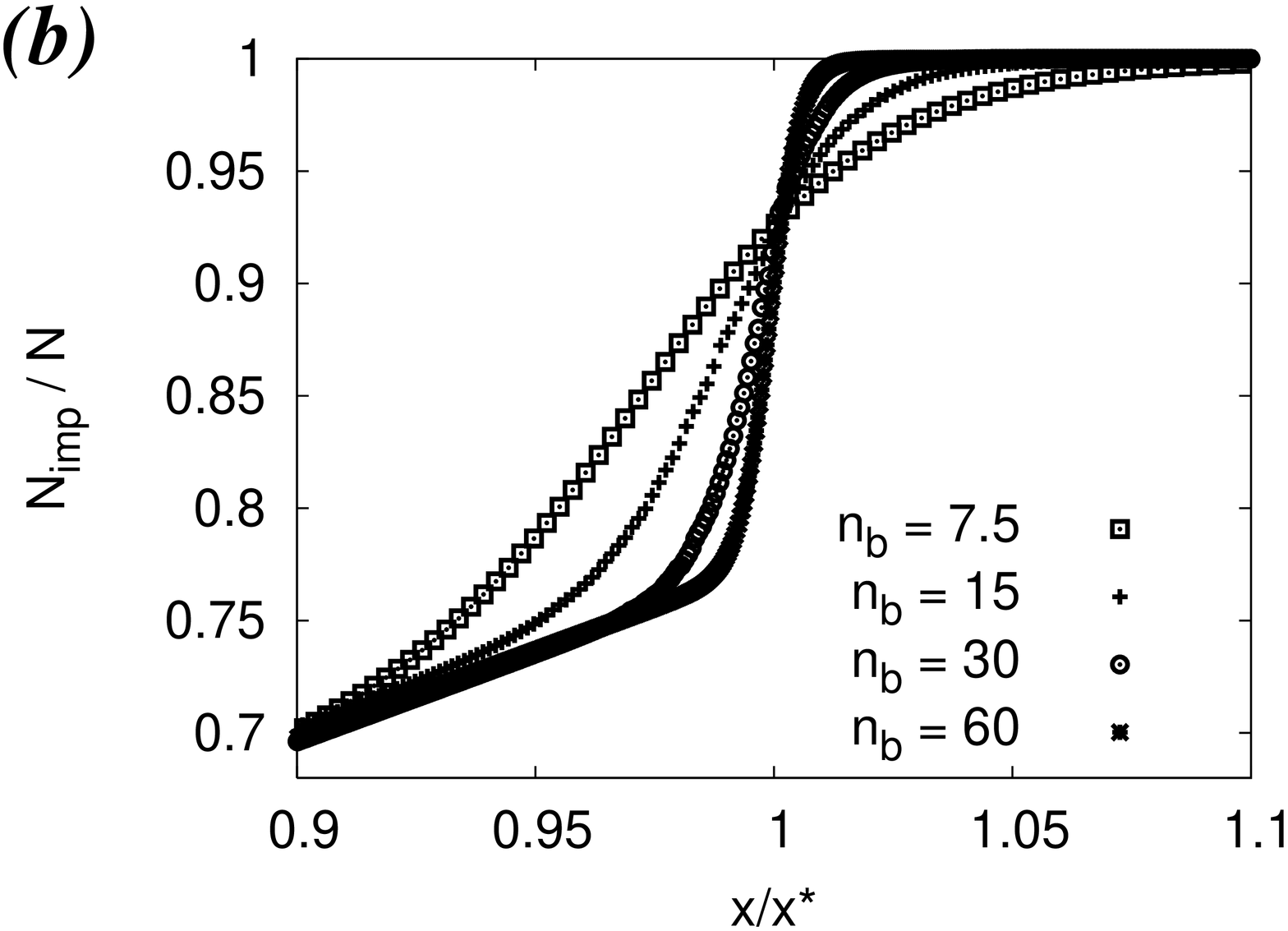}
\caption{(a) Average fraction of imprisoned units, $N_{\rm imp}/N$,
plotted against the reduced end coordinate $x/x^*$ for $n_b=60$
and for various values of $D$.
The solid line gives the theoretical prediction
that the fraction increases linearly with $x$ below the transition point,
and it is in perfect agreement with our data.
At the transition point $x/x^*=1$, $N_{\rm imp}/N$ jumps up from
$0.76$ to $1$, and the relative reduction in the number of imprisoned monomers
is then $\Delta_N \approx 0.24$. (b) $N_{\rm imp}/N$ vs. $x/x^*$
the transition region, $x/x^*=1$, for $D=21$ and for different
number of $n_b$, displaying the rounding of the transition
is due to the finiteness of the number of blobs.}
\label{drag-fig7}
\end{center}
\end{figure}

\begin{figure}
\begin{center}
\includegraphics[scale=0.31,angle=270]{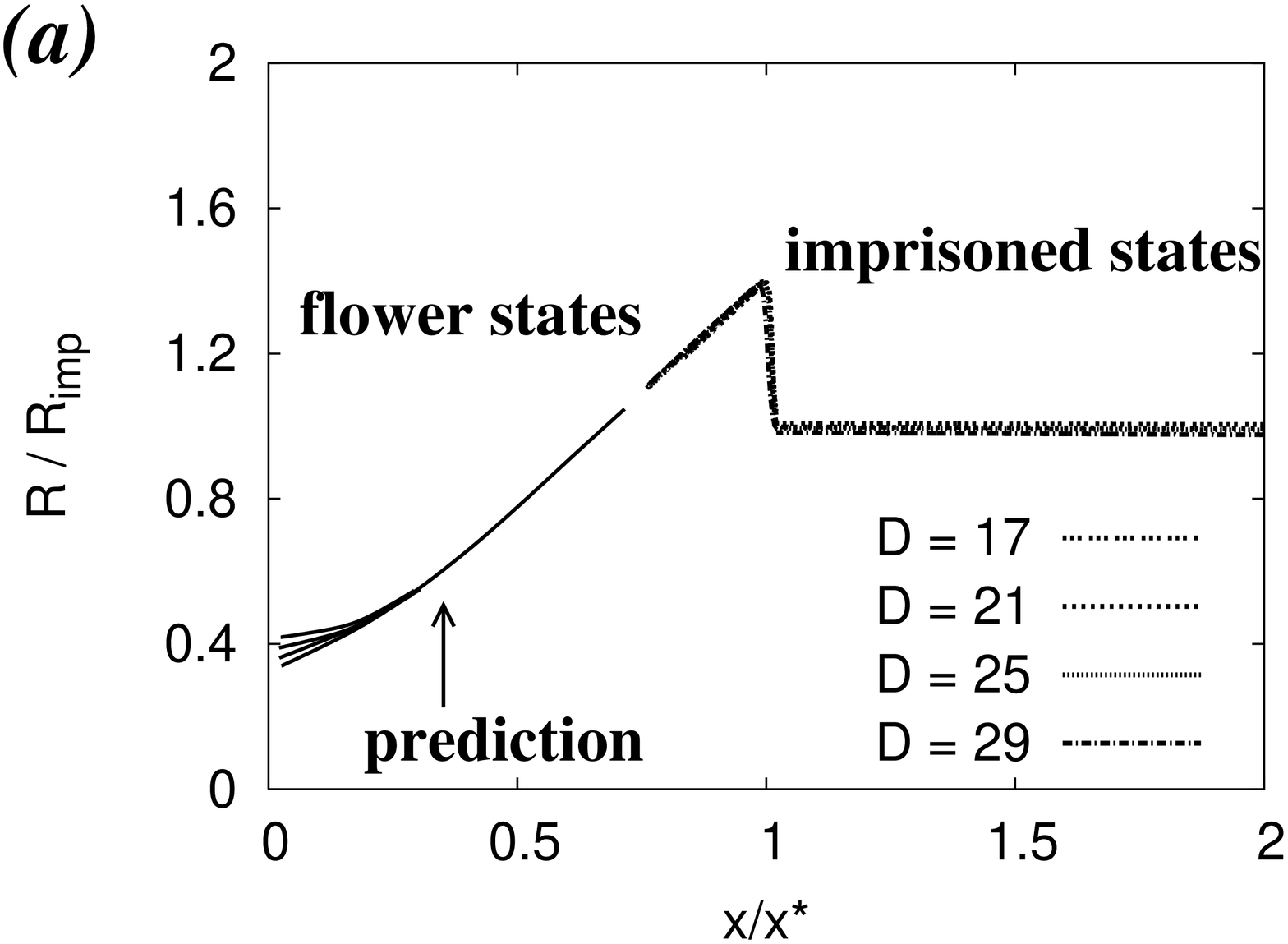}\hspace{0.4cm}
\includegraphics[scale=0.31,angle=270]{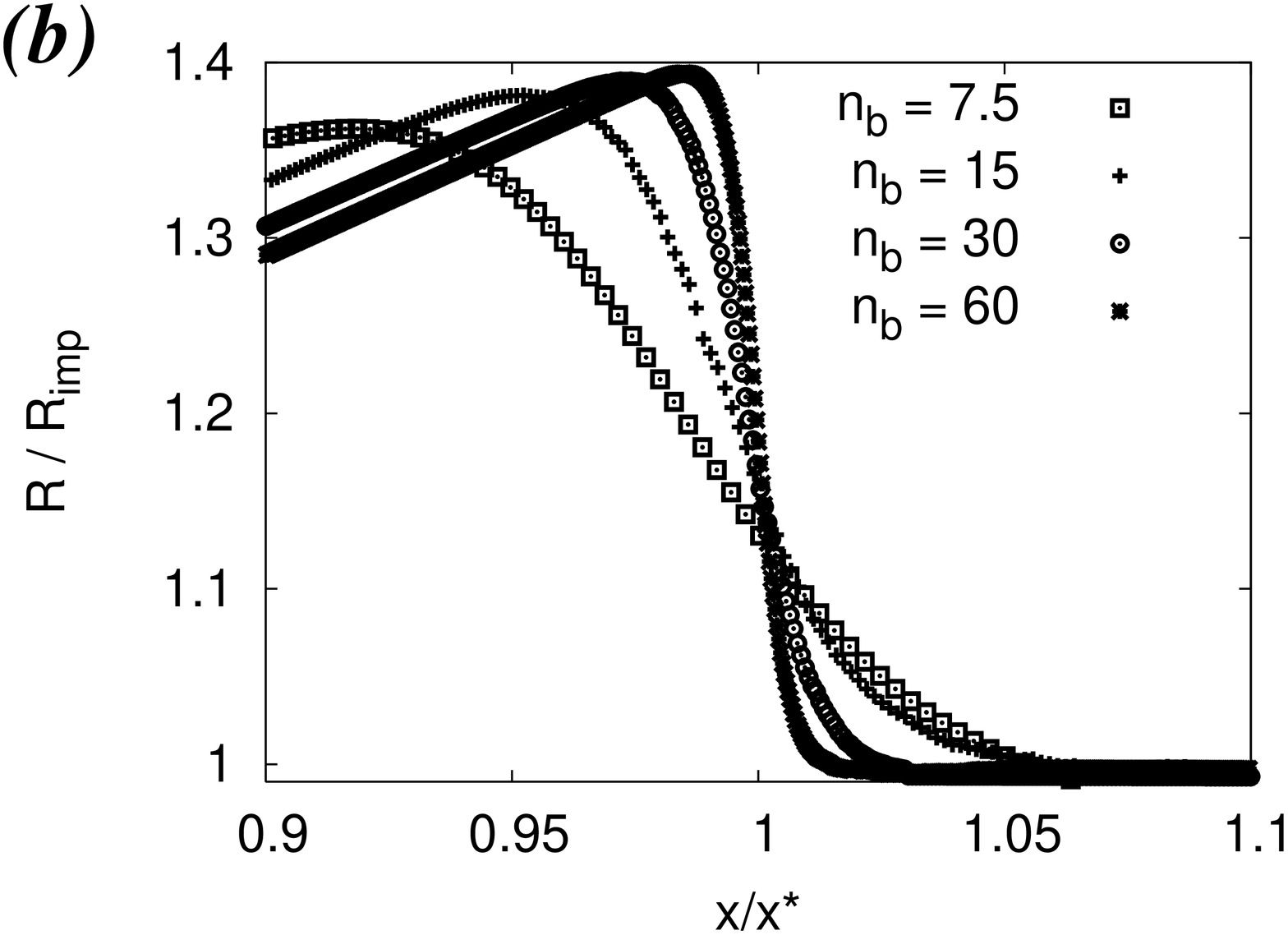}
\caption{(a) Reduced rms end-to-end distance $R/R_{\rm imp}$,
plotted against the reduced end coordinate $x/x^*$
for $n_b=60$ and for various values of $D$.
$R_{\rm imp}$ is the average rms end-to-end distance for a fully
confined chain. Near $x=0$, chains behave as a self-avoiding coil,
$R/R_{\rm imp}$ depends on both $N$ and $D$.
Approaching the transition point from below, a linear behavior
appears according to the scaling prediction.
At the transition point $x/x^*=1$, the value
of $R/R_{\rm imp}$ jumps down from $1.38$ to $1$ and the reduced jump of
end-to-end distance $\Delta_R=(1.38-1)/1.38 \approx 0.27 $.
(b) $R/R_{\rm imp}$ vs. $x/x^*$ around the transition region, $x/x^*=1$,
for $D=21$ and for different number of $n_b$.}
\label{drag-fig8}
\end{center}
\end{figure}

\begin{figure}
\begin{center}
\includegraphics[scale=0.31,angle=270]{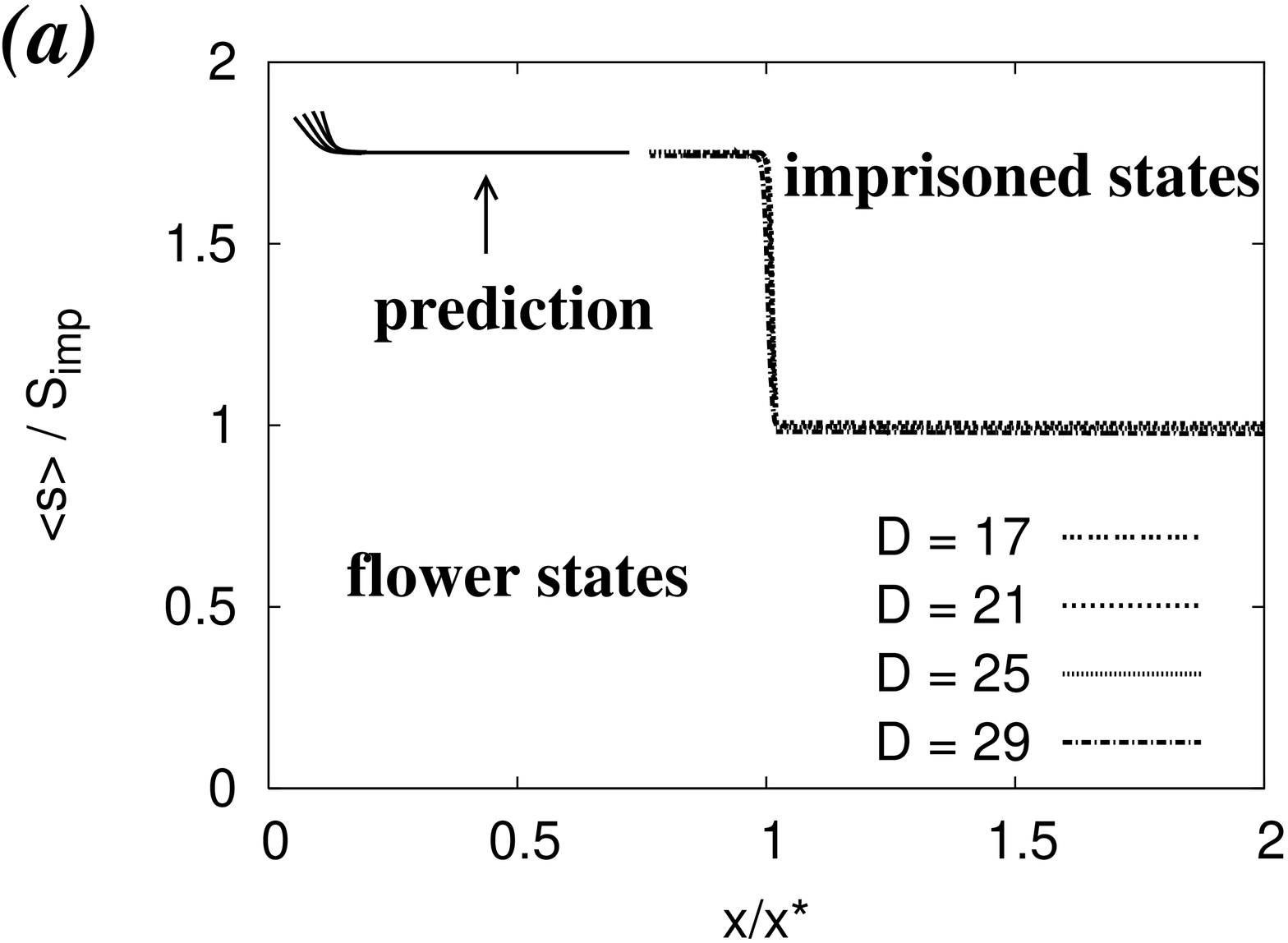}\hspace{0.4cm}
\includegraphics[scale=0.31,angle=270]{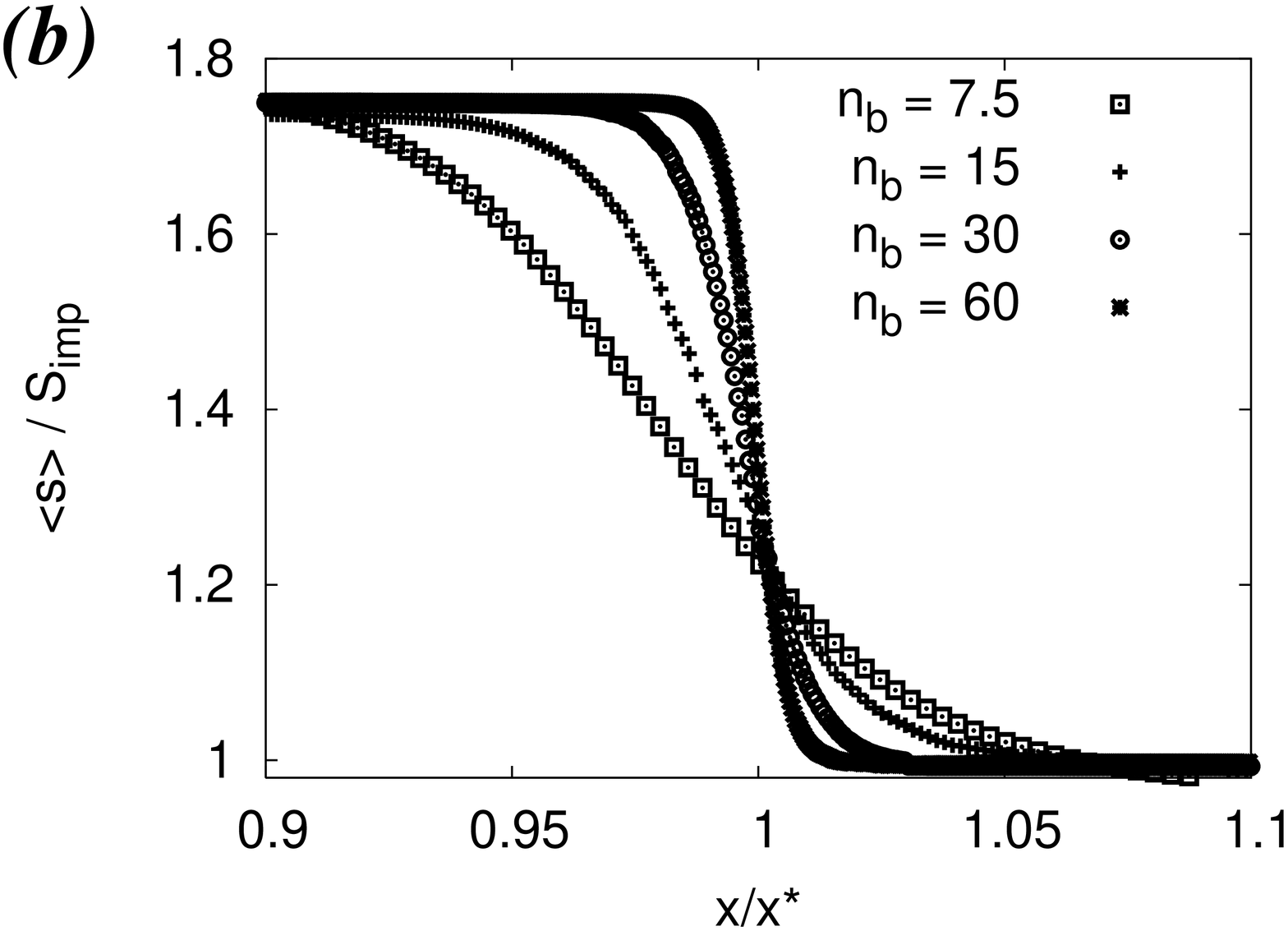}
\caption{(a) Reduced average stretching parameter $<s>/S_{\rm imp}$, plotted
against the reduced end coordinate $x/x^*$ for $n_b=60$ and for
various values of $D$. $S_{\rm imp}$ is the average
stretching of a fully confined chain, and it does not depend on $x/x^*$.
The average stretching of the chain in a flower state is the stretching of
the stem and it is also independent of $x/x^*$ except the region that
near $x=0$ where it depends
on both $N$ and $D$. At the transition point $x/x^*=1$,
the value  $<s>/S_{\rm imp}$ jumps down from $1.75$ to $1$ and the
reduced jump of the stretching
parameter is $\Delta_S=(1.75-1)/1.75 \approx 0.43$.
(b) $<s>/S_{\rm imp}$ vs. $x/x^*$ around the transition region, $x/x^*=1$,
for $D=21$ and for different number of blobs $n_b$.}
\label{drag-fig9}
\end{center}
\end{figure}

The intersection of the two free energy branches defines the transition
point, or the critical distance $x^*$ away from the open end of the tube.
Equating the two expressions of eq~\ref{Fscaling} one obtains:
\be
    x^*=(B_{\rm imp}/B_{\rm fl}) N(D/a)^{1-1/\nu} = 1.26(4)N(D/a)^{1-1/\nu}
\label{transition point}
\ee
MC data for the transition points displayed in Figure~\ref{drag-fig6}
are in full agreement with eq~\ref{transition point}.

The average fraction of imprisoned units, $N_{\rm imp}/N$, as a function of the
reduced end coordinate $x/x^*$ is shown in Figure~\ref{drag-fig7}. It is
clear that this fraction increases linearly with $x$ and jumps up to $1$ at the
transition point. This jump represents the abrupt ``slurping'' of the tail
which constitutes approximately $1/4$ of the total number of units, $N$.
As mentioned above, truly sharp jumps can occur in the thermodynamic
limit $(N \rightarrow \infty)$ only, so when one looks at the behavior of
$N_{\rm imp}/N$ with high resolution on the scale $x/x^*$ for
various sizes of the system, one can
resolve the finite-size rounding (Figure~\ref{drag-fig7}b).
The next graph (Figure~\ref{drag-fig8}) represents the change in the rms
end-to-end distance reduced by the corresponding value for the completely
confined chain, $R/R_{\rm imp}$.
The starting value at $x=0$ represents an unconfined coil with $R\approx N^\nu$
so that the reduced value depends on both $N$ and $D$, i.e., $R/R_{\rm imp}
\propto N^\nu/(ND^{1-1/\nu})$. Then, a linear dependence appears which is
terminated at the transition point. The average size jumps down by
approximately
$25\%$.
Finally, the average stretching parameter $<s>$  of the confined
part of the chain is shown in Figure~\ref{drag-fig9} as a function
of the same reduced distance $x/x^*$.
(For the precise definition of the stretching parameter $s$, see the next section).
It is reduced by the value $S_{\rm imp}$
characterizing the fully confined chain, independent of $x$ and given by
eq~\ref{Simprison}. In a flower state the stretching parameter $S_{\rm fl}$
describes the average stretching of the stem. It is independent of $x$ also and
is about $1.75$ times larger than that for the fully confined chain. A small
deviation exists near $x=0$ only. Beyond the transition point $S_{\rm fl}$ jumps
down to the $S_{\rm imp}$ value.

\section{Landau theory} 

Unlike the case of critical phenomena, the Landau theory approach 
is very well suited for analyzing first-order 
transitions, including possible metastable states. 
The idea is to  first subdivide all configurations into subsets 
associated with a given value of an appropriately chosen order 
parameter $s$ that allows to distinguish between different states 
or phases. Landau free energy $\Phi(s)$ is the free energy of 
a given subset, and is therefore a function of the order parameter. 
The minimum value of the Landau free energy is attained for the subset 
that contains most of the equilibrium configurations, and therefore 
coincides with the equilibrium free energy of the system, $F$.
Far enough from the first order transition point, the Landau free
energy has only one minimum. However, near the transition the function 
is expected to have two minima (the deeper one is stable and the other 
is metastable). Exactly at the transition point, 
both minima are of equal depth.
We define the order parameter as the \underline{chain stretching} in the fully confined
state, $s=r/(N_{\rm imp} a)$ ($N_{\rm imp}=N$) where $r$ is the instantaneous end-to-end 
distance of the chain, or as \underline{the stretching of the stem} only in the 
flower state, $s=x/(na)$ where $n$ is the instantaneous number 
of confined monomers in the stem, and $x$ is the length of the stem.
The Landau function consists of two branches that 
have to be introduced separately. For fully confined configurations the 
Landau free energy, up to an additive constant, is directly 
expressed in terms of the distribution of the end-to-end distance 
$P(r|N,D)$ of a chain in the tube:
\be
   \Phi_{\rm imp}(r/Na) = const-\ln P(r|N,D) \; .\label{DefinitionF(r)}
\ee
There exists no closed formula for such a distribution of 
confined chains with excluded volume interactions. 
In Ref.~\cite{Sotta} the (non-normalized) distribution for the gyration radius was 
studied numerically and the following scaling form was proposed:
\be
   \ln P(R_{g}|N,D) = - N(D/a)^{-1/\nu}A \left[ u^{-\alpha} + 
Bu^{\delta}\right]  \label{P(Rg)}
\ee
where $\alpha =(3\nu -1)^{-1}$ , $\delta =(1-\nu)^{-1}$ and
$u={(R_{g}/Na)}(D/a)^{{-1+1/\nu}}$.
The parameters $A$ and $B$ are non-universal numbers of order unity 
and they do not depend on $N$ or $D$.
The first term describes the concentration effects in the 
des Cloizeaux~\cite{Cloizeaux} form, and the second term is the Pincus~\cite{Pincus} 
scaling form of the stretching free energy.
From our MC simulations, we have found that the end-to-end distribution
$P(r|N,D)$ and 
the equilibrium free energy $F$ are well described by a similar 
scaling formula corrected by an additional $r$-independent term, namely
the Landau free energy for the imprisoned state is then
\be
   \Phi_{\rm imp}(s) = N(D/a)^{-1/\nu}A \left[ u^{-\alpha} + B u^{\delta}
+C\right] \label{Pimp}
\ee
where $u$ is now related to our order parameter $s$
\be
u={(r/Na)} (D/a)^{{-1+1/\nu}} = s (D/a)^{-1+{1/\nu}} \; . \label{uimp} 
\ee

  This branch is limited to the range of values for the order parameter, $\thinspace$
$0<s<x/Na$. $\enspace$ In the thermodynamic limit, the average value of the order  parameter of a
fully confined chain, $S_{\rm imp}=<s>$, is found by locating the minimum 
of $\Phi_{\rm imp}(s)$, i.e., $d\Phi_{\rm imp}(s)/ds=0$ at $s=S_{\rm imp}$,
and the corresponding minimum of the Landau free energy is  
the equilibrium free energy $F$. Using eq~\ref{uimp}, we obtain the equilibrium average 
value of the end-to-end distance $R_{\rm imp}=<r>$,  
\be
       R_{\rm imp}=N<s>=N (D/a)^{1-{1/\nu}} u_{1}\label{R equilibr}
\ee 
where $u_{1}=(\alpha/\delta B)^{1/(\alpha+\delta)}$ gives the position of the minimum 
of the function $f(u)=u^{-\alpha} + B u^{\delta} +C$.
Comparing this result with the MC data of $D=17$, i.e. eq~\ref{Rimprison} 
with taking $A_{\rm imp}=0.94$,
we immediately determine the numerical value of $B=0.67$.
Using eq~\ref{Pimp}, the equilibrium free energy $F_{\rm imp}$ is the Landau free 
energy at $u=u_1$:
\be
F_{\rm imp}=N(D/a)^{-1/\nu}A \left(1.67 +C\right) \label{Fimp equilibr}
\ee
Comparing this again with the MC data of $D=17$, 
i.e. eq~\ref{Fimprison} with taking $B_{\rm imp}=5.38$,
we get a relationship between the coefficients $A$ and $C$. 
The last condition that eventually fixes all the numerical coefficients 
$A$, $B$ and $C$ of the Landau function is obtained by analyzing its second branch.

   For the partially confined chains in the flower state, since in fact only 
$n$ monomers that comprise the stem of the flower give contributions to the 
free energy, instead of $N$ and $r$, we use $n$ and $x$ in eq~\ref{Pimp}. The formula 
of the Landau free energy is therefore,  
\ba
  \Phi_{\rm fl}(s)&=&\frac{x}{D}A \left[ u^{-\alpha-1} + B u^{\delta-1}
+Cu^{-1}\right],
\enspace s \geq \frac{x}{N}  \;.     \label{Pfl}
\ea
where $u$ is given by
\be
   u={(x/na)} (D/a)^{{-1+1/\nu}} = s (D/a)^{-1+{1/\nu}} \;.
\ee
The average value of the order parameter $S_{\rm fl}$, and the 
equilibrium free energy $F_{\rm fl}$ for the flower state are obtained 
by the same procedure as that for the imprisoned state. 
Demanding that $S_{\rm fl}$ and the coefficient with the $x/D$ 
factor in eq~\ref{Pfl} both coincide with the MC results, 
we are uniquely fixing the numerical values of parameters
$A=1.48$, $B=0.67$ and $C=1.98$. 

\begin{figure}
\begin{center}
\includegraphics[scale=0.31,angle=270]{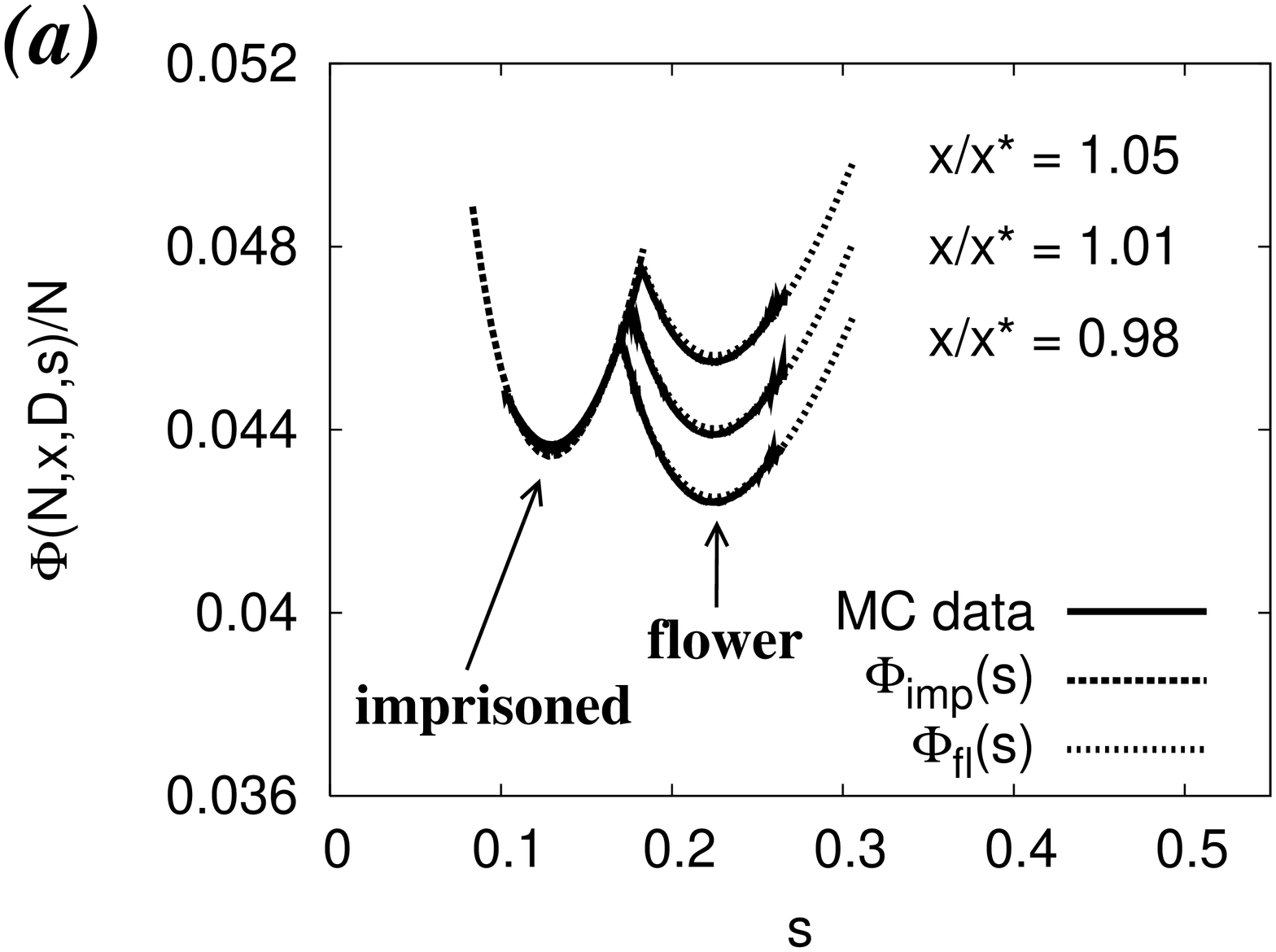}\hspace{0.4cm}
\includegraphics[scale=0.31,angle=270]{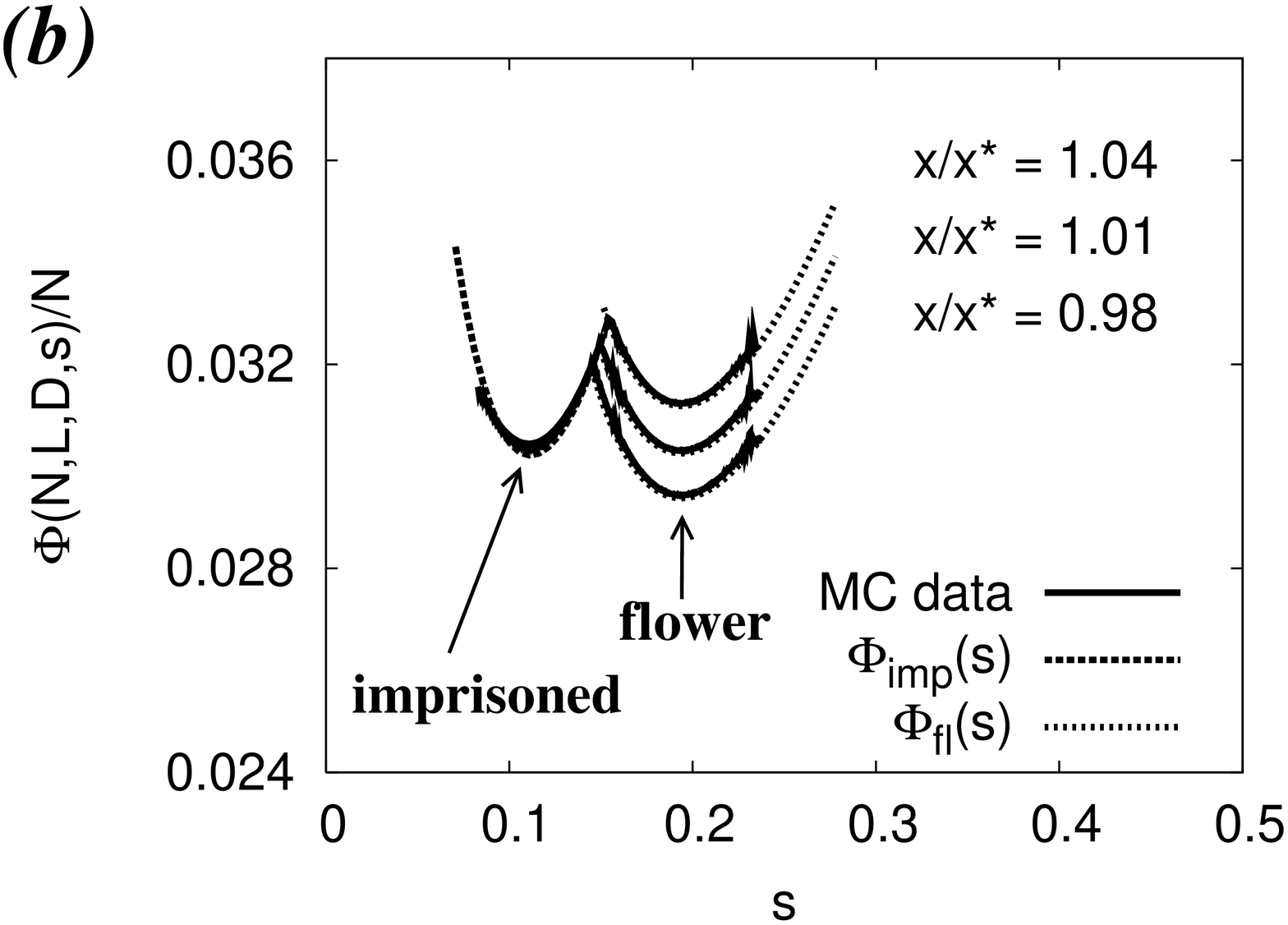}
\caption{Landau free energy divided by $N$, $\Phi(N,x,D,s)/N$,
plotted against the order parameter $s$ for various values of $x/x^*$ and for
$D=17$ (a) and $D=21$ (b). The predicted Landau free energy
function $\Phi_{\rm imp}$, eq~\ref{Pimp}, in the imprisoned
regime, and $\Phi_{\rm fl}$, eq~\ref{Pfl}, in the flower
regime, with $A=1.48$, $B=0.67$, and $C=0.98$,
give a good fit to the MC data.}
\label{drag-fig10}
\end{center}
\end{figure}

\section{Comparison with MC results.}

For our simulations, the Landau free energy as a function of $s$,
$\Phi(N,x,D,s)$, is given by taking the logarithm of
the properly normalized accumulated histogram of stretching parameter $s$.
Results for tubes with diameters $D=17$ and $D=21$ near the transition
point $x^*/Na \sim 1.26(D/a)^{1-1/\nu}$ \{eq~\ref{transition point}\}
are shown in Figure~\ref{drag-fig10}.
Two branches of the analytical Landau function given by 
eqs~\ref{Pimp} and \ref{Pfl} for $A=1.48 $, $B=0.67$, and $C=1.98$ 
are also presented in Figure~\ref{drag-fig10}.  
We see that in both the flower and the imprisoned states, 
MC data are in perfect agreement with
the theoretical predictions for a wide range of $s$ at any $x$, and $D$ 
although the coefficients $A$, $B$, and $C$
are determined by using only the MC estimates of average stretching 
parameters $S_{\rm imp}$ and $S_{\rm fl}$
, and of average free energies $F_{\rm imp}$
and $F_{\rm fl}$ for $D=17$.  

From the analytical Landau function in the imprisoned state and
in the flower state, eqs~\ref{Pimp} and \ref{Pfl}, and
the determined values of the coefficients $A$, $B$, and $C$,
we obtain the following scaling relationships of  
the equilibrium order parameters $s$ 
\be
s=s_{eq} = \left\{
\begin{array}{ll}
 0.94(D/a)^{1-1/\nu}=S_{\rm imp} & \enspace {\rm imprisoned \, state} \\
 1.64(D/a)^{1-1/\nu}=S_{\rm fl}  &\enspace {\rm flower \, state}\\
\end{array} \right . \; ,
\label{S eqlb}
\ee
and the equilibrium free energies $F$,
\be
F_{eq} = \left\{
\begin{array}{ll}
  5.40N(D/a)^{-1/\nu}=F_{\rm imp} & \enspace {\rm imprisoned \, state} \\
 4.27x/D=F_{\rm fl}  &\enspace {\rm flower \, state}\\
\end{array} \right . 
\label{F eqlb}
\ee
The transition point is found from the condition that the two minima of 
the Landau free energy function are of equal depth. 
Using eqs~\ref{F eqlb} we get
\be
  \frac{x_{tr}}{Na} \sim 1.26(D/a)^{1-1/\nu}
\label{tr point}
\ee

The reduced jump of the order parameter is therefore
\be
 \Delta_S=\frac{S_{\rm fl}-S_{\rm imp}}{S_{\rm fl}}
\approx 0.43 \;.  \label{jumpS}
\ee
The average number of units dragged into the tube for an imprisoned 
state is $N_{\rm imp}=N$, while for the coexisting flower state
we have only $N_{\rm imp}=<n>=x/S_{\rm fl}$ monomers. 
Using eqs~\ref{S eqlb} and \ref{tr point} we obtain the 
relative reduction in the number of imprisoned monomers
\be
   {\Delta_N}=\frac{N-x/S_{\rm fl}}{N} \approx 0.23 \;.
  \label{jumpN}
\ee
Finally, the reduced jump of the end-to-end distance is obtained by 
using eqs~\ref{S eqlb} and \ref{tr point},  
\be
   {\Delta_R}=\frac{x-R_{\rm imp}}{x} \approx 0.25 \;.
\label{jumpR}
\ee
Equations~\ref{jumpS}-\ref{jumpR} show that the sizes
of jumps in $S$, $N_{\rm imp}/N$, and $R$ are universal
quantities (i.e., independent of $D$ and $N$), and they satisfy the following relation 
\be
   {(1-\Delta_R)}{(1-\Delta_N)}={(1-\Delta_S)} \;.
\label{jump connection}
\ee
Comparing with the MC results shown in Figures~\ref{drag-fig7}-\ref{drag-fig9},
we see that the Landau theory gives a good qualitative and quantitative
agreement.

\begin{figure}
\begin{center}
\includegraphics[scale=0.31,angle=270]{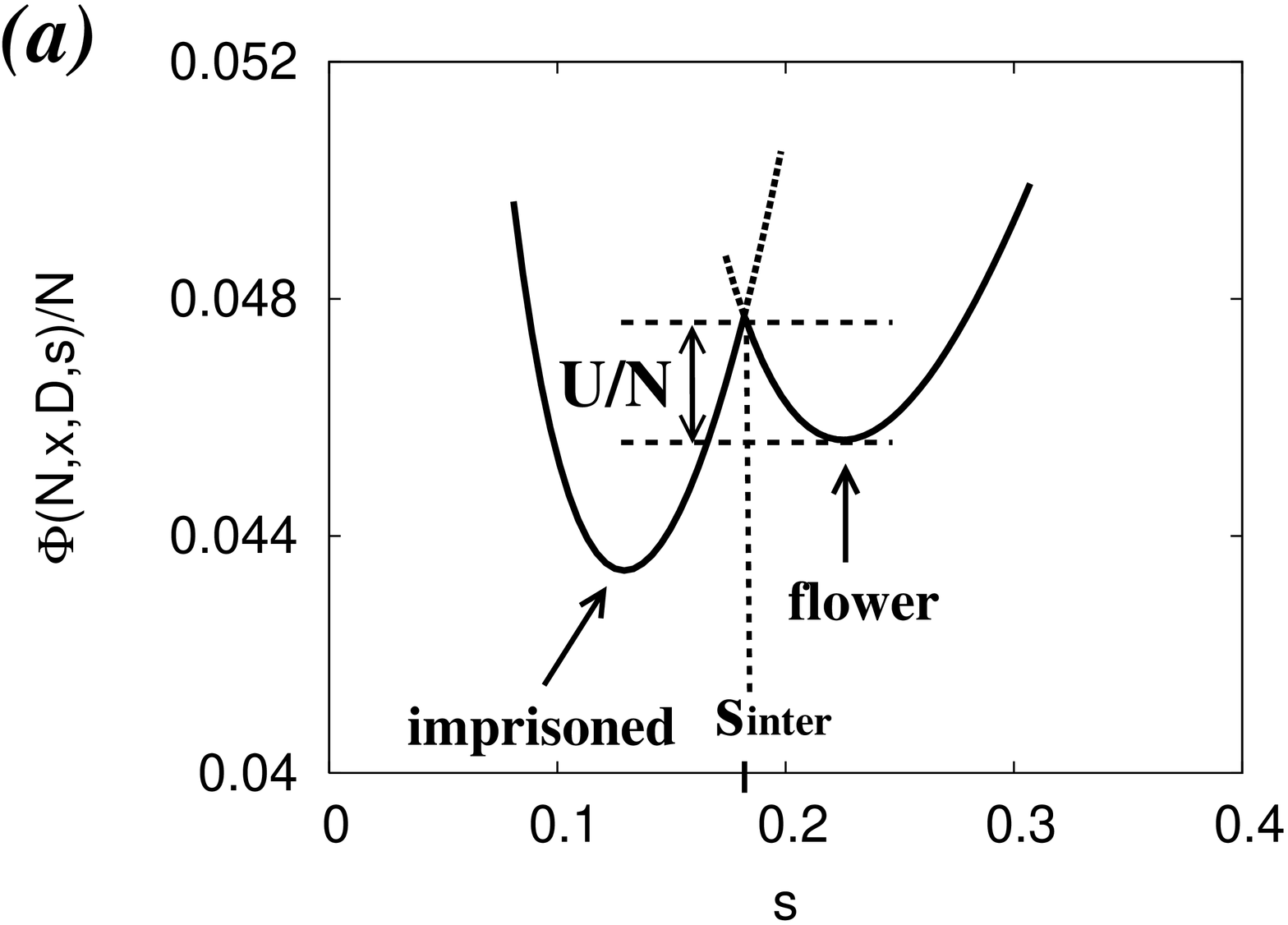}\hspace{0.4cm}
\includegraphics[scale=0.31,angle=270]{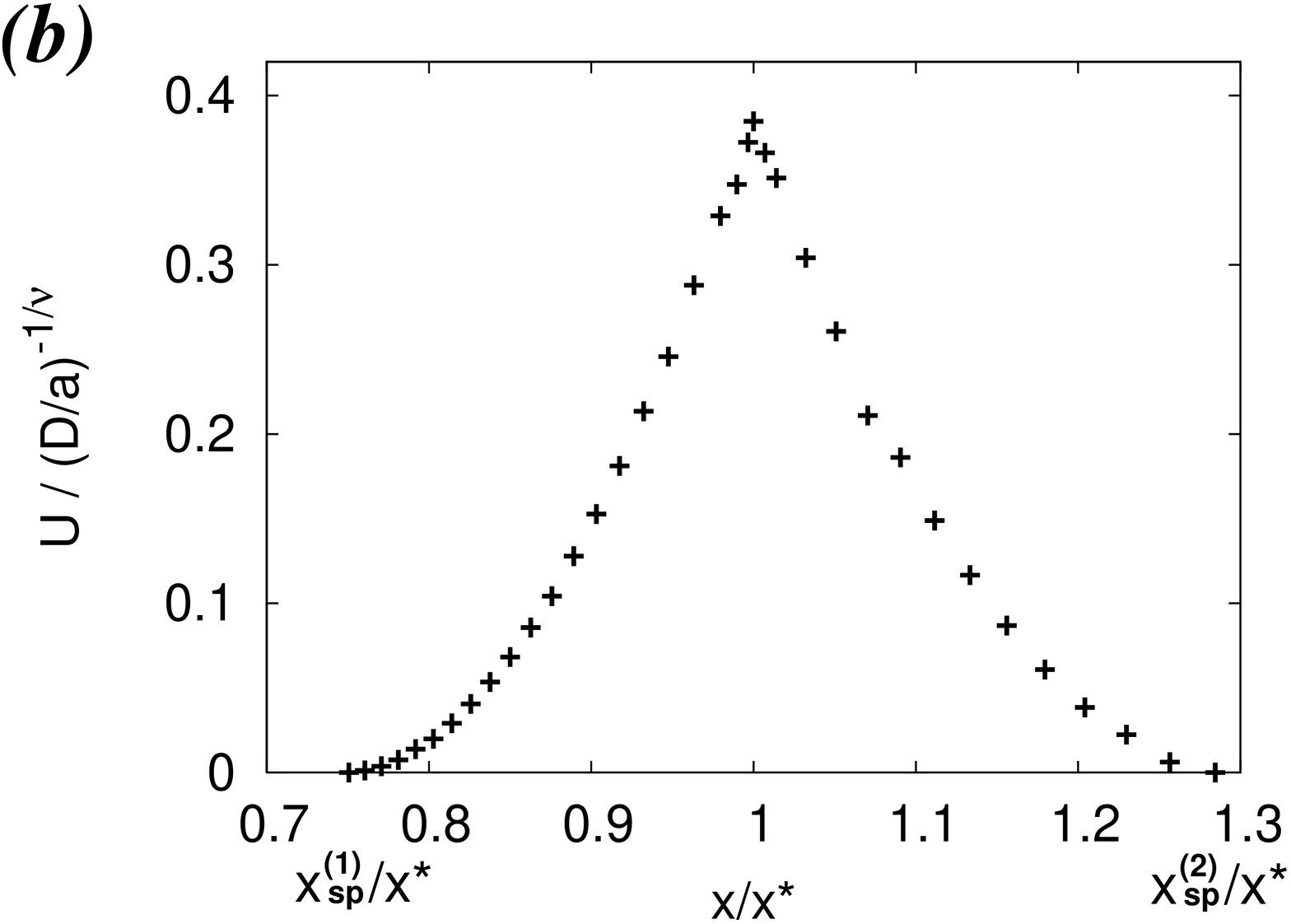}
\caption{(a) Landau free energy per segment, $\Phi(N,x,D,s)/N$,
versus order parameter $s$ close to the transition point for $x/x^* = 1.05$
and for $D=17$.
The two branches intersect at some intermediate state with $s=s_{\rm inter}$.
From the difference between
$\Phi(N,x,D,s)/N$ at the intersection point and
at the higher minimum one finds $U/N =0.0021 $.
(b)Barrier heights per blob, $U/n_b$, estimated as described in (a), vs. $x/x^*$.
 In the coordinates used the curve is universal (independent on $D$, and $N$)
At the transition point $x/x^* =1$ the barrier height is maximal.
At the two spinodal points $x_{\rm sp}^{\rm (1)}/x^*=0.78$ and $x_{\rm
sp}^{(2)}/x^*=1.33$ the barrier vanishes.}
\label{drag-fig11}
\end{center}
\end{figure}

\begin{figure}
\begin{center}
\includegraphics[scale=0.31,angle=270]{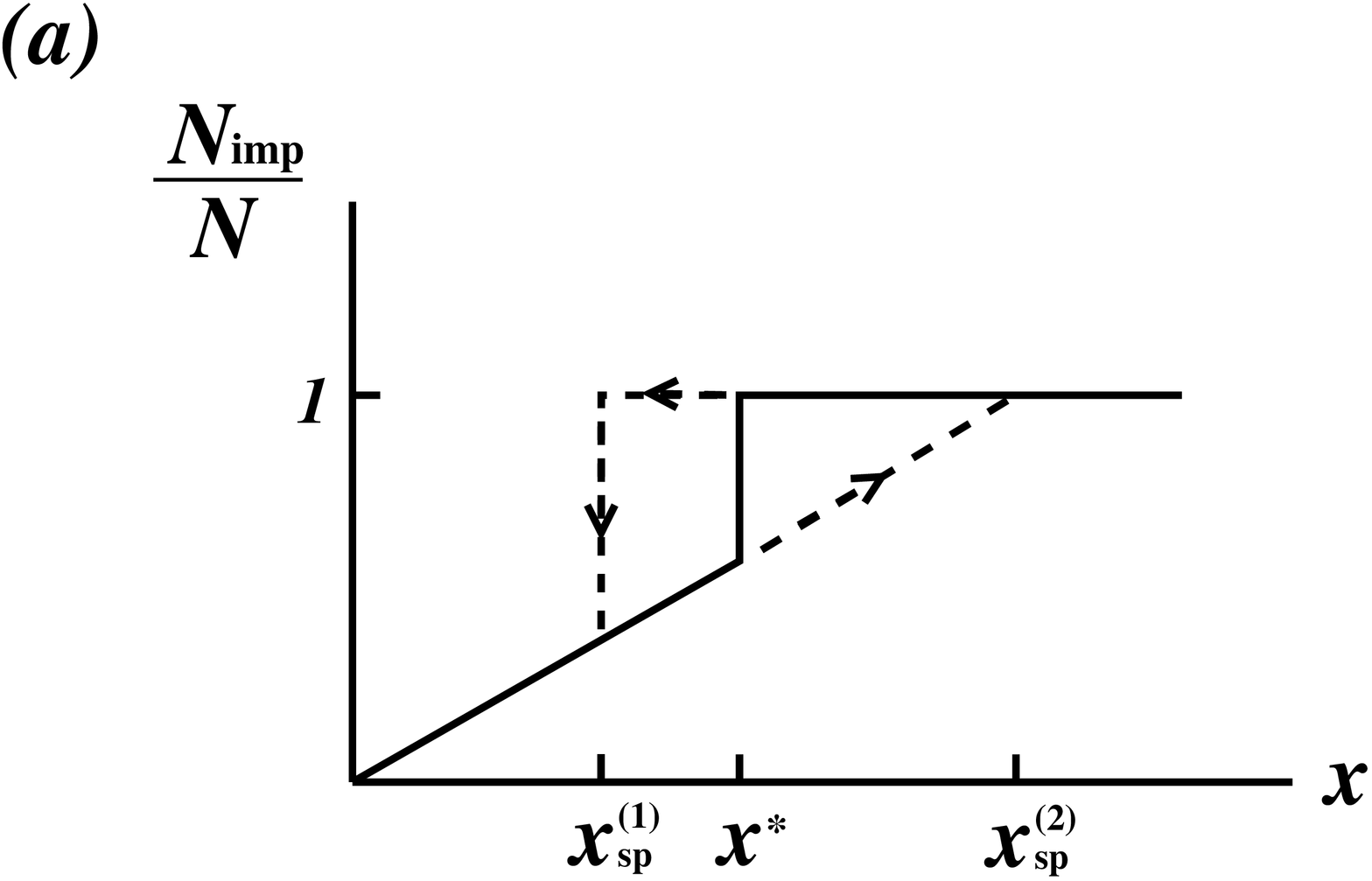}\hspace{0.4cm}
\includegraphics[scale=0.31,angle=270]{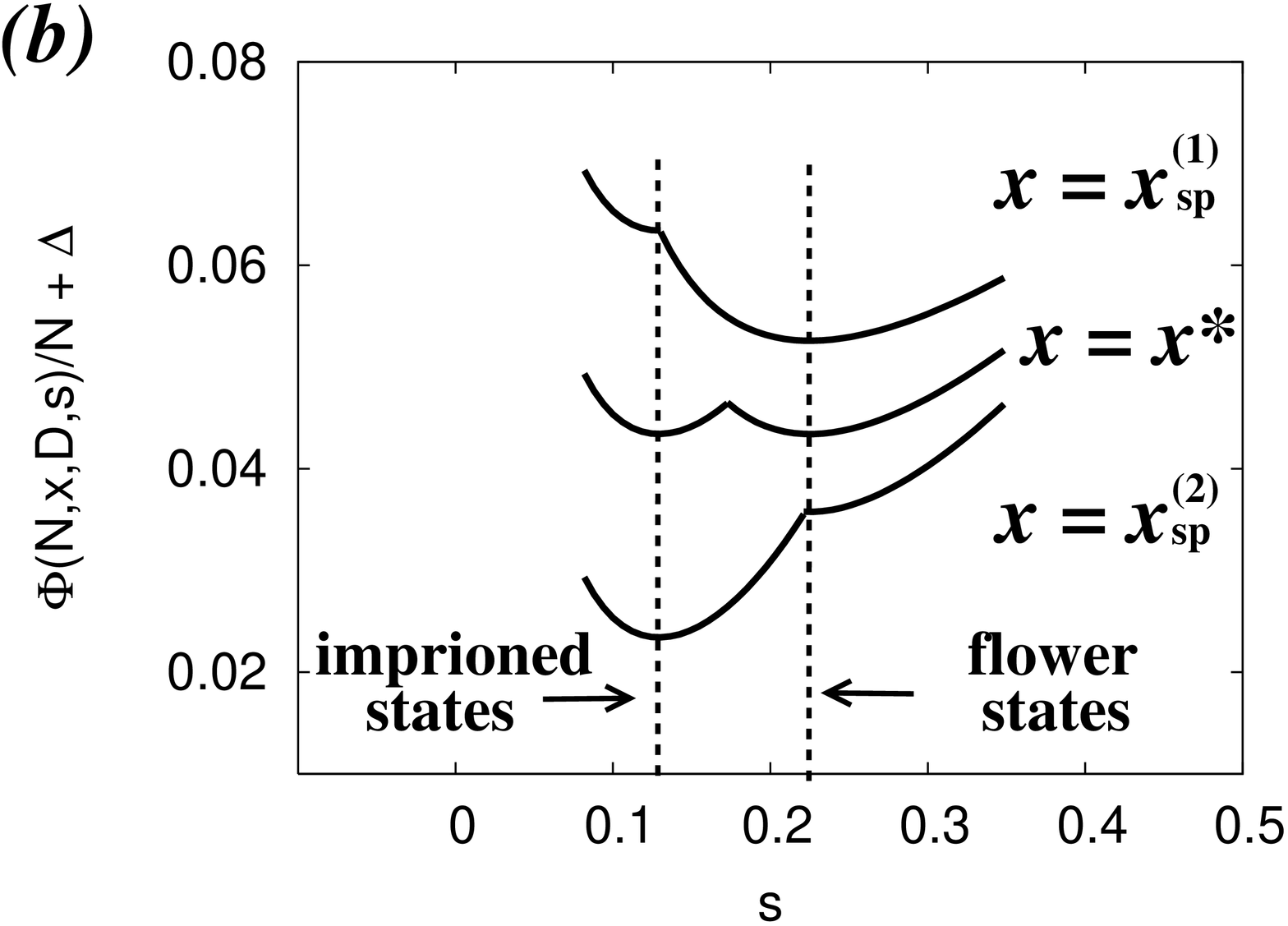}
\caption{(a) Number of imprisoned units (normalized to
unity), $N_{\rm imp}/N$, plotted against the coordinate of the end
monomer position of a dragged chain, $x$. If the chain end is moved into
the tube very slowly, $N_{\rm imp}/N$ of chains in equilibrium states
follows the solid curve.
If the chain end is moved quickly, the chain may be trapped in the metastable
states (dashed curves) and a hysteresis loop appears.
(b) Landau free energy divided by $N$, $\Phi(N,x,D,s)/N$,
as a function of the order parameter $s$ for $D=17$ at the
transition point $x^*$
and at the two spinodal points $x_{\rm sp}^{\rm (1)}$
and $x_{\rm sp}^{\rm (2)}$. In order to distinguish these three curves well,
a constant value $\Delta$ is added to $\Phi(N,x,D,s)/N$.
$\Delta=0.02$, $0$, and $-0.02$ for $x=x_{\rm sp}^{\rm (1)}$, $x^*$,
and $x=x_{\rm sp}^{\rm (2)}$, respectively.}
\label{drag-fig12}
\end{center}
\end{figure}

\section{Metastable regions and spinodal points}

When the chain is dragged into the tube by one end slowly (quasi-statically),
the number of imprisoned units, $N_{\rm imp}$ as a
function of $x$ grows linearly up to the
transition point $x^*$ and then jumps to $N$ corresponding to a full confinement
as shown in Figure~\ref{drag-fig7}. This
would mean that at any value of $x$ a complete equilibrium is achieved and only
the lowest minimum of the Landau free energy is populated.
However, Figure~\ref{drag-fig11}a shows clearly that for $x > x^*$ there still
exists another minimum of the Landau free energy representing the
metastable flower state. The barrier height per blob turns out to be a
universal function of the reduced coordinate $x/x^*$ that is displayed in
Figure~\ref{drag-fig11}b.
So, if the chain end is
moved into the tube relatively quickly compared to the metastable lifetime, the
chain will be trapped in the flower state, and the number of imprisoned units
will keep increasing linearly as shown by the dashed line in
Figure~\ref{drag-fig12}a. As the tail decreases, so does the barrier
height, as demonstrated in Figure~\ref{drag-fig11}b until
the metastability is completely lost at a spinodal point $x=x_{\rm sp}^{(2)}$.
The spinodal value $x_{\rm sp}^{\rm (2)}$ is defined by the
condition that the flower minimum coincides with the matching point
(Figure~\ref{drag-fig12}b) leading to
\be
   x_{\rm sp}^{\rm (2)}=Na S_{\rm fl}
\label{spinodal1}
\ee
where $S_{\rm fl}$ is given by eq~\ref{S eqlb}.

If the chain is in a fully confined state and its end is moved back to the tube
entrance quasi-statically, the equilibrium curve is retraced. On the other
hand, a metastable imprisoned state also appears at $x < x^*$.  Its local
stability
is lost at the other spinodal point,
\be
   x_{\rm sp}^{\rm (1)}=Na S_{\rm imp}
\label{spinodal2}
\ee
where $S_{\rm imp}$ is again given by eq~\ref{S eqlb}. It follows from
the physical meaning of $S_{\rm imp}$ that $x_{\rm sp}^{(1)}$ coincides with the
equilibrium end-to-end distance for a fully confined chain.
Note that both spinodal points scale in the same way with $N$ and $D$,
$x_{\rm sp}\sim ND^{1-1/\nu}$ although with different numerical prefactors. The
hysteresis loop associated with the metastable states is displayed in Figure
\ref{drag-fig12}a. 
Let us estimate the lifetime of a metastable state of a single $\lambda$ phage
DNA confined in a nanochannel~\cite{Reisner} with the following parameters:
contour length $L=16\mu m$, persistence length $a=50 nm$, tube diameter $D=150
nm$. This gives the number of blobs $(L/a)(D/a)^{-1/\nu}=50$. The lifetime of a
metastable state (mean first passage time) can be estimated as
$\tau_{ms}=\tau_0\exp(U/k_BT)$ where $U$ is the height of the barrier separating
the metastable minima and the interaction point.  
A characteristic relaxation time $\tau_0$
for the DNA molecule estimated from autocorrelated extension fluctuations 
is close to 1 sec. 
The barrier height near the transition point is about $0.38$ $k_BT$
per blob according to Figure~\ref{drag-fig11}b. With these parameters an
estimate for the lifetime of a metastable state is astronomical 
, $\tau_{ms} \sim 10^{13}s$, leaving no chance to observe the 
equilibrium transition experimentally and
making hysteresis effects inevitable. However, since the number of blobs depends
strongly on the width of the tube one can expect that experiments with a wider
tube would be much closer to equilibrium. As an example, for the same DNA
molecule in a tube with $D=500 nm$ a similar estimate gives $n_b=10$ and
$\tau_{ms} \sim 1 min$.

\section{Escape of a released chain}

We are now in a position to address the second part of the problem announced 
in the title of the paper. We have demonstrated that dragging a chain into a tube by 
its end involves a first order transition accompanied by a jump-wise change in the 
chain conformation. The question is then, whether its escape back from the tube upon 
release will also involve a jumpwise transition. To clarify the situation we recall that 
we were trying to simulate and describe theoretically an experimental situation where 
the position of one chain end serves as a parameter controlled by external means, e.g. 
by using optical tweezers. Theoretically, this implies a statistical description in a 
constant $x$ ensemble in which statistical averaging is done over all internal degrees of 
freedom at fixed values of $x$ (together with other parameters such as $N$, $D$, and 
temperature $T$). The averaged quantities, e.g. the average number of imprisoned units, 
are thus functions of $x$. Their variations with $x$ describe the response of thermally 
fluctuating variables to a change in external parameters.

\begin{figure}
\begin{center}
\epsfig{file=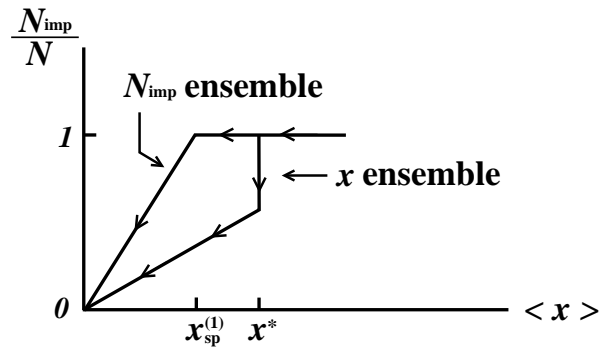, width=4.5cm, angle=270}
\caption{The number of imprisoned units (normalized to unity), 
$N_{\rm imp}/N$, plotted against
the end monomer position of the released chain, $x$, in two ensembles.
In $x$-ensemble, the end position changes quasi-statically. 
In $N_{\rm imp}$-ensemble,
the number of imprisoned units changes quasi-statically.}
\label{drag-fig13}
\end{center}
\end{figure}

In a gradual escape of a released chain initially confined in a tube the 
coordinates of all segments, including both ends, are themselves subject to thermal 
fluctuations. An important question to be addressed is: what is the appropriate 
statistical ensemble to describe this process? 
Let us first assume that the position of the distant end inside the tube, $x$ ,
is indeed a dynamic variable that changes much slower than the other degrees of
freedom. (We argue below that this assumption is generally incorrect unless
specific mechanisms are in place to achieve this effect). Then the evolution of
$x$ itself will be governed by the free energy profile $F(x)$ discussed earlier
and shown in Figure~\ref{drag-fig5}. Dynamically, the chain will diffuse
along the flat horizontal portion of the slope and slide down the slope towards
the de-confined state at $x < 0$. In the process, 
both the average $x$, $<x>$, and the
average $n$, $N_{\rm imp}$, will be changing with time but $N_{\rm imp}$ as a 
function of $<x>$ defined
parametrically will follow the retraced equilibrium curve in
Figure~\ref{drag-fig13}.

Now the relaxation time for the number of imprisoned units, $N_{\rm imp}$, presumed to be
governed by faster dynamics is to be estimated. For $x$ in the range between two
spinodal points $x_{\rm sp}^{\rm (1)} < x < x_{\rm sp}^{\rm (2)}$, there are two minima separated
by a barrier, see Figure~\ref{drag-fig12}b leading to a very slow relaxation
that involves barrier crossing. It is clear that this contradicts the assumption
that $N_{\rm imp}$ adjusts quickly to any change in $x$ unless the motion in the $x$
coordinate is specifically slowed down by some additional mechanism. As an
example of such a mechanism, one could envisage a situation where the distant
chain end is modified to have a sticky anchor attached. This would not affect the
fluctuations of the end nearest to the tube opening and thus would not slow down
the process of equilibrating the number of imprisoned segments that involves
expulsion of a tail. A relatively large colloidal particle attached to the
distant chain end for the purpose of using optical tweezers may produce a
similar effect. 

A more appropriate ensemble seems to be the one where the number of imprisoned
segments is assumed to change quasi-statically while the end position $x$
is adjusted by thermal fluctuations. In the absence of any external force the
relaxation of $x$ is purely diffusive if the chain far inside the tube with
$N_{\rm imp}=N$. Once the process of gradual escape starts and $N_{\rm imp}<N$ 
the coordinate $x$ relaxes to produce the equilibrium stretching of the remaining 
confined part at the value $s=S_{\rm imp}$. This relaxation is never controlled by barrier
crossing rate. The evolution of $<N_{\rm imp}>(t)$ itself can be pictured as a slide down
the linear slope of the free energy $F(N_{\rm imp}) = N_{\rm imp} B_{\rm imp}(D/a)^{-1/\nu}$
towards the minimum $n=0$. The parametric dependence of $<N_{\rm imp}>$ vs. $<x>$ in
this process is also shown in Figure~\ref{drag-fig13}. 

The use of the fixed $N_{\rm imp}$ ensemble would be clearly justified if the 
evolution of the variable $N_{\rm imp}$ was to be explicitly slowed down 
without affecting the relaxation rate of the distant end. 
An experimental situation where this slowing down could be realized if 
the chain was escaping from the tube trough a 
partially blocked opening, similar to a setting normally assumed in 
a problem of translocation through a narrow hole in a thick membrane. 

It follows from the above discussion that the behavior of a released chain
escaping from a tube is a problem of real polymer dynamics which is far from 
being
well understood. By using quasi-equilibrium statistical ensembles we were able
to clarify the two limiting dynamic cases when one of the global variables is
much slower than the other. Although both limiting results may be applicable to
real experimental situation provided some additional modifications of the basic
setting are introduced, one could speculate that the process of an escape from a
tube in the simplest setting is somewhere in between these limits.
A more general and powerful approach where both $N_{\rm imp}$ and $x$
are treated dynamically is sketched in the Appendix.

\section{Conclusions}
In this paper, the statistical mechanics of a long flexible
polymer dragged into a cylindrical tube with repulsive wall
under good solvent conditions is studied, both via a scaling
theory and by Monte Carlo simulations, using the PERM algorithm
that allows the successful study of very long chains.
It is shown that in the limit of infinite chain length
an entropically driven abrupt transition occurs when the distance
of the chain end from the tube entrance, $x$, is used as a 
control parameter. Experimentally, this situation could
be realized e.g. when a nanoparticle is attached to
this chain end and the position of this particle is controlled
externally by a laser tweezer.
This has to be distinguished from the case when the chain is driven
by applying a constant force (e.g. electrophoretically).
It is shown that a critical value $x_c$ exists, such that
for $x<x_c$ a finite fraction of monomers is outside of the
tube in a mushroom-like configuration, and the remaining
$N_{\rm imp}<N$ monomers form the ''stem", a one-dimensional
stretched string of blobs, while for $x>x_c$ the ''crown"
of this flower-like conformation of the polymer has disappeared,
and all $N$ monomers have been sucked into the tube to become
part of the ''stem", the string of blobs. This transition
for $N \rightarrow \infty$ is strongly discontinuous, since at the 
transition $N_{\rm imp}/N$ jumps from about $3/4$ to unity in our model.
We construct a suitable order parameter for this transition and
use scaling ideas to formulate the Landau free energy
branches of both states that compete near the transition with
each other. The Monte Carlo results confirm the general
picture of these two phases and allow to estimate the undetermined
prefactors of the Landau theory description. The Monte Carlo results also
allow to quantify the extent of finite size rounding of the transition
that inevitably occurs as a consequence of the finiteness of the
chain length. Arguments are presented that for cases of physical interest
(such as DNA in artificial nanopores) it is rather likely that
this transition is affected by hysteresis, and estimates 
for the lifetime of metastable states are given.

  Interestingly, no transition is predicted for a reverse process of
chains release if the constraint fixing
the chain end at a particular position $x$ is removed,
so that no external force ever acts at the chain end inside the tube:
the chain then can reduce its free energy continuously
by "escaping out" of the tube. The dynamics of this chain
expulsion process from a tube is an interesting problem for
further study, however.

   Another interesting extension would concern the behavior of
a chain dragged into a tube with attractive walls. We expect
that such a situation could be of interest in the context of 
polymer translocation through membranes.

\section*{Acknowledgments}
We are grateful to the Deutsche Forschungsgemeinschaft (DFG)
for financial support: A.M.S. and L.I.K. were supported under grant 
Nos. 436 RUS 113/863/0, and H.-P.H. was supported under grant NO SFB
625/A3. A.M.S. received partial support under grant NWO-RFBR 047.017.026, 
and RFBR 08-03-00402-a. Stimulating discussions with A. Grosberg and
J.-U. Sommer are acknowledged.

\section*{Appendix: A dynamic picture of polymer escape from a tube}

We have demonstrated that descriptions of the process of polymer chain de-confinement
upon release may differ depending on the statistical ensemble. The choice of
an appropriate ensemble depends, in turn, on which of the variable: the distant
end position, $x$, or the number of imprisoned segments, $N_{\rm imp}$, is the 
slowest. In a more consistent approach, both variables are treated on equal basis.
We define the Landau free energy as a function of two independent variables,
$\Phi(x,N_{\rm imp})$ (assuming, as usual, that all other internal degrees of
freedom equilibrate much quicker). We have all the necessary information to
define $\Phi(x,N_{\rm imp})$ at hand. For a fully confined chain with 
$N_{\rm imp}=N$ and far enough inside the tube, $x>x_{\rm sp}^{\rm (1)}$,
the chain is fully relaxed and its free energy is given by $F_{\rm imp}$,
see eq~\ref{Fimp equilibr}. For a fully confined chain with $x<x_{\rm sp}^{(1)}$,
variable $x$ has the meaning of the end-to-end distance and the expression 
derived for the flower branch of the Landau free energy, eq~\ref{Pfl},
applies. The same expression applies for any $x$ provided the chain is only 
partially confined. This eventually gives

\be
\frac{\Phi(x,N_{\rm imp})}{n_b} =  \left\{
\begin{array}{l}
  5.40,  \qquad {\rm for \,} N_{\rm imp}=N\,, \, x \ge 0.94Dn_b \\
  1.48 \frac{x}{Dn_b} \left(u^{-\alpha-1}+0.67 u^{\delta -1} + 1.98 u^{-1}\right), \\
\qquad \qquad \qquad {\rm otherwise}\\
\end{array} \right .
\label{phidyn}
\ee
Here $n_b=N(D/a)^{-1/\nu}$ is the number of blobs in a fully confined chain, and $u$
is the ratio of two reduced quantities $u={(x/(Dn_b))}/{(N_{\rm imp}/N)}$.

\begin{figure}
\begin{center}
\epsfig{file=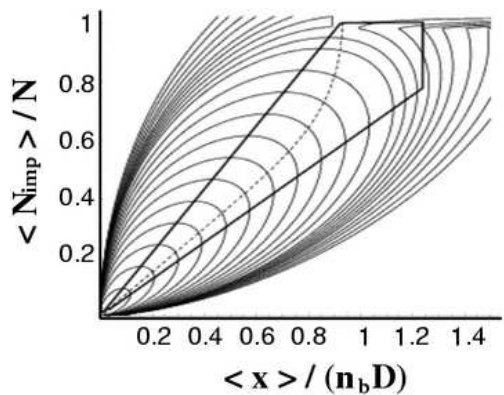, width=5.5cm, angle=270}
\caption{Contour plot of Landau free energy landscape as a function of two
independent variables $\Phi(x,N_{\rm imp})$. The corridor of the fully confined
relaxed chain in the upper right corner is artificially broadened for a clearer
visual picture. Also shown are the parametric trajectories
$<N_{\rm imp}(t)>$ vs. $<x(t)>$ of the de-confinement process for two limiting
cases $D_x << D_n$ and $D_x >> D_n$ (solid lines) and for isotropic diffusion
$D_x=D_n$ (dashed line).}
\label{drag-fig14}
\end{center}
\end{figure}

A contour plot showing the landscape of $\Phi(x,N_{\rm imp})$ is presented
in Figure~\ref{drag-fig14}. States corresponding to a fully confined relaxed chain
are depicted by a narrow corridor in the upper right part of the landscape.
The lower left corner represents a completely free chain with the lowest 
possible free energy. The corridor is generally separated from the sloping
landscape by a barrier, except for an opening in the vicinity of $x=x_{\rm sp}^{(1)}$.

Full dynamic evolution in the $(x,N_{\rm imp})$ configuration space will be 
governed by a Fokker-Planck equation with Landau free energy playing the role
of effective potential. However, if one avoids the detailed description
of barrier-crossing events, the problem is simplified considerably.
The dynamics of averaged quantities $<x(t)>$ and $<N_{\rm imp}(t)>$ is
then driven by thermodynamic forces $\frac{\partial \Phi(x,N_{\rm imp})}{\partial x}$
and $\frac{\partial \Phi(x,N_{\rm imp})}{\partial N_{\rm imp}}$, respectively,
and the coupled non-linear equations of motion are given as
\be
 \frac{d<x(t)>}{dt} = -D_x \frac{\partial \Phi}{\partial x}
\ee
and
\be
 \frac{d<N_{\rm imp}(t)>}{dt} = -D_n \frac{\partial \Phi}{\partial N_{\rm imp}}
\ee
where $D_x$ and $D_n$ are the effective diffusion coefficients along the $x$ and
$N_{\rm imp}$ coordinates, correspondingly. Geometrically, this evolution
is a diffusive slide along the slopes of the Landau free energy in a two-dimensional
configuration space. A detailed discussion of the diffusion coefficients and 
the physics behind them is well beyond the scope of this paper.

Here we briefly present the result of a dynamic analysis for three different
scenarios. First we analyze the two limiting cases discussed in Section 8.
Equations of motion were solved numerically for very slow $x$ dynamics with
$D_x = 10^{-3} D_n$ and the curve of $<N_{\rm imp}(t)>$ vs. $<x(t)>$
parametrically defined by $t$ is shown in Figure~\ref{drag-fig14}. It is clear
that this coincides with the quasi-static trajectory predicted in the $x$-ensemble. 
Another curve was obtained from a numerical solution assuming slow $N_{\rm imp}$
dynamics with $D_n=10^{-3}D_x$. This coincides precisely with the 
$N_{\rm imp}$-ensemble description as evidenced by comparing Figures~\ref{drag-fig13}
and~\ref{drag-fig14}. Finally, the parametric curve for the case of isotropic diffusion 
$D_x = D_n$ is also presented. This is characterized by a relatively rapid initial
growth of the ejected part of the chain, the distant end position evolution
catching up with some delay. Eventually, the trajectory slide down the valley along
its geometrical bottom line. The shapes of all three trajectories are 
insensitive to assumptions made about the change in diffusion coefficients during
the de-confinement process as long as their ratio is kept fixed.


\begin{thebibliography}{99}

\bibitem{Salman} Salman, H.; Zbaida, D.; Rabin, Y.; Chatenay, D.;
Elbaum, M. {\sl Proc. Natl. Acad. Sci. U.S.A.} {\bf 2001}, {\sl 98}, 7247.

\bibitem{Meller} Meller, A. {\sl J. Phys.: Condens. Matter} {\bf 2003}, 
{\sl 15}, R581.

\bibitem{Kasianowicz} Kasianowicz, J. J.; Brandin, E.; Branton, D.; 
Deaner, D. W.
{\sl Proc. Natl. Acad. Sci. U.S.A.} {\bf 1996}, {\sl 93}, 13770.

\bibitem{Aktson} Aktson, M.; Branton, D.; Kasianowicz, J. J.; Brandin, E.;
Deaner, D. W. {\it Biophys. J.} {\bf 1999}, {\sl 77}, 3227. 

\bibitem{Meller00} Meller, A.; Nivon, L.; Brandin, E.; Golovchenko, J. A.;
Branton, D. {\sl Proc. Natl. Acad. Sci. U.S.A.} {\bf 2000}, {\sl 97}, 1079.

\bibitem{Clausen} Clausen-Schaumann, H.; Seitz, M.; Krautbauer R.; 
Gaub, H. E. {\sl Curr. Opin. Chem. Biol.} {\bf 2001}, {\sl 4}, 524.

\bibitem{Williams} Williams, M. C.; Rouzina, I.
{\sl Curr. Opin. Struct. Biol.} {\bf 2002}, {\sl 12}, 330.

\bibitem{Reisner} Reisner, W.; Morton, K. J.; Riehn, R.; Wang, Y. M.;
Yu, Z.; Rosen, M.; Sturm, J. C.; Chou, S. Y.; Frey, E.; Austin, R. H.
{\sl Phys. Rev. Lett.} {\bf 2005}, {\sl 94}, 196101.

\bibitem{g97} Grassberger, P. {\sl Phys. Rev. E} {\bf 1997}, {\sl 56}, 3682.

\bibitem{Hsu03} Hsu, H.-P.; Grassberger, P. {\sl Eur. Phys. J.}
{\bf 2003}, {\sl B36}, 209.

\bibitem{Hsu04} Hsu, H.-P.; Grassberger, P. {\sl J. Chem. Phys.}
{\bf 2004}, {\sl 120}, 2034.

\bibitem{Hsu07} Hsu, H.-P.; Binder, K.; Skvortsov, A. M.; Klushin, L. I.
{\sl Phys. Rev. E} {\bf 2007}, {\sl 76} 021108.

\bibitem{Sotta} Sotta, P.; Lesne, A.; Victor, J. M.
{\sl J. Chem. Phys.} {\bf 2000}, {\sl 112}, 1565.

\bibitem{Kremer} Kremer, K.; Binder, K. {\sl J. Chem. Phys.}
{\bf 1984}, {\sl 81}, 6381.

\bibitem{Milchev94} Milchev, A.; Paul, W.; Binder, K.
{\sl Macromol. Theory Simul.} {\bf 1994}, {\sl 3}, 305.

\bibitem{Yang} Yang, Y.; Burkhardt, T. M.; Gompper, G. 
{\sl Phys. Rev. E} {\bf 2007}, {\sl 76}, 011804.

\bibitem{Daoud} Daoud, M.; de Gennes, P. G. {\sl J. Phys. (paris)}
{\bf 1977}, {\sl 38}, 85.

\bibitem{Randel} Randel, R.; Loebl, H.; Matthai, C. 
{\sl Macromol. Theory Simul.} {\bf 2004}, {\sl 13}, 387.

\bibitem{Subra} Subramanian, G.; Williams, D. R. M.; Pincus, P. A.
{\sl Europhys. Lett.} {\bf 1995}, {\sl 29}, 285; 
{\sl Macromolecules} {\bf 1996}, {\sl 29}, 4045.

\bibitem{Ennis99} Ennis, J.; Sevick, E. M.; Williams, D. R. M.
{\sl Phys. Rev. E} {\bf 1999}, {\sl 60}, 6906.

\bibitem{Sevick99} Sevick, E. M.; Williams, D. R. M.
{\sl Macromolecules} {\bf 1999}, {\sl 32}, 6841.

\bibitem{Steels} Steels, B. M.; Leermakers, F. A. M.; Haynes, C. A.
{\sl J. Chrom. B.} {\bf 2000}, {\sl 743}, 31.

\bibitem{Milchev} Milchev, A.; Yamakov, V.; Binder, K.
{\sl Phys. Chem. Chem. Phys.} {\bf 1999}, {\sl 1}, 2083;
{\sl Europhys. Lett.} {\bf 1999}, {\sl 47}, 675.

\bibitem{Cloizeaux} des Cloizeaux, J. {\sl J. Phys. (Paris)} 
{\bf 1980}, {\sl 41}, 223.

\bibitem{Pincus} Pincus, P. {\sl Macromolecules} {\bf 1976}, {\sl 9}, 386.

\bibitem{Dimitar} Dimitrov, D. I.; Milchev, A.; Binder, K.; Klushin, L. I.; 
Skvortsov, A. M. {\sl e-print} arXiv:0802.3116v1 {\bf 2008}.

\end{thebibliography}
\end{document}